# Polaritonic Quantum Matter

D.N. Basov[1], A. Asenjo-Garcia[1], P.J. Schuck[1], X.-Y. Zhu[1], A. Rubio[2,3], A. Cavalleri[2], M. Delor[1], M.M. Fogler[4], Mengkun Liu[5,6]

1. Columbia University, New York, New York 10027
2. Max Planck Institute for the Structure and Dynamics of Matter and Center for Free-Electron Laser Science, Luruper Chaussee 149, 22761 Hamburg, Germany
3. Center for Computational Quantum Physics (CCQ) and Initiative for Computational Catalysis (ICC),The Flatiron Institute, 162 Fifth Avenue, New York NY 10010
4. Department of Physics, University of California, San Diego, CA 92093
5. Department of Physics and Astronomy, Stony Brook University, Stony Brook, NY 11794
6. National Synchrotron Light Source II, Brookhaven National Laboratory, Upton, New York 11973

**Abstract.** Polaritons are quantum mechanical superpositions of photon states with elementary excitations in molecules and solids. The light-matter admixture causes a characteristic frequency-momentum dispersion shared by all polaritons irrespective of the microscopic nature of material excitations that could entail charge, spin, lattice or orbital effects. Polaritons retain the strong nonlinearities of their matter component and simultaneously inherit ray-like propagation of light. Polaritons prompt new properties, enable new opportunities for spectroscopy/imaging, empower quantum simulations and give rise to new forms of synthetic quantum matter.  Here, we review the emergent effects rooted in polaritonic quasiparticles in a wide variety of their physical implementations. We present a broad portfolio of the physical platforms and phenomena of what we term *polaritonic quantum matter*. We discuss the unifying aspects of polaritons across different platforms and physical implementations and focus on recent developments in: polaritonic imaging, cavity electrodynamics and cavity materials engineering, topology and nonlinearities, as well as quantum polaritonics.

# SECTION 1: Meet the polariton

Light and matter can hybridize to yield composite quasiparticles referred to as polaritons. The notion of dual light-matter polaritons was introduced by Hopfield in a seminal paper on the optics of dielectrics in 1958[1]. Since then, polaritons have been extensively studied in many experimental settings, ranging from quantum materials[2] to cavity and waveguide quantum electrodynamics (QED) with atoms[3], molecules[4], solid-state emitters and superconducting qubits[5–7,] (Fig.1). An alphabetized summary of polaritonic effects in atomic, molecular and solid state systems in offered in Ref.[2]. Because the term "polariton" is extensively used across distinct subfields, many separate but related ideas and phenomena are categorized under that same notion. In this brief review, we expound on the notion of polaritons as a unifying concept connecting subfields of contemporary quantum many-body physics and quantum information technology. Each of these subfields has produced a vast literature; thus, omissions in our narrative, though regrettable, are all but inevitable.

## 1A. Polaritons as hybrid light-matter states

Polaritons are bosonic quasiparticles that embody the normal modes of the interacting photon-matter systems. They can be viewed either as excitations of a medium coupled by electromagnetic forces or, equivalently. as photons that are dressed by interaction with the medium. In the latter case, polaritons can be thought of as heavy photons with a renormalized dispersion characterized by a reduced group velocity and/or a nonzero effective mass. The polariton group velocity describes ray-like propagation of polariton wavepackets (Fig. 1 and Fig. 2C). Polariton dispersion typically contains two or more branches split by avoided crossing(s).

*Formal definitions and Hopfield coefficients.* Polaritons can be seen as linear superpositions $u|g\rangle|1\rangle + v|e\rangle|0\rangle$ such that $|g\rangle$ and $|e\rangle$ are, respectively, the ground and excited states of the medium, $|0\rangle$ and $|1\rangle$ are the states with zero and one photons, and $u$ and $v$ are the so-called Hopfield coefficients [1] normalized by the condition $|u|^2 + |v|^2 = 1$. The relative weights of the "light" and "matter" terms are therefore given by $|u|^2$ and $|v|^2$. The term "polariton" is sometimes utilized to describe only the immediate vicinity of the avoided crossing region (shaded area in Fig. 1B) where these weights are comparable. In this formulation, one also requires the light-matter coupling to be "strong;" that is, it must fulfill the condition that the gap separating the dispersion branches (commonly referred to as the vacuum Rabi frequency $\Omega_R$) exceeds the combined line broadening (Box 1). The photon-matter content of polaritons is strongly skewed towards one or the other side at the extremes of the polaritonic dispersion. However, many aspects of polariton physics are observable even when either the light or the matter component is only minute, as in the case, for example, of surface plasmon polaritons in metals[8]. Furthermore, as long as the dielectric permittivity of the medium remains large, referring to its normal modes as polaritons is fully justified[9]. Polaritons in cavities may involve 0,1,2,3…,n photons, with the zero-photon state being the vacuum state of a cavity (Box 2 and Section 2B).

*On the many varieties of polaritons.* Polaritons in quantum materials (QMs) can be formed by the hybridization of light with a variety of dipole-active excitations such as phonons, excitons, plasmons, magnons, Cooper pairs, etc. (Fig. 1B; see also Box 2, highlighting theoretical aspects of states' dressing). In the context of atomic, molecular, and optical (AMO) physics, polaritons are optically dressed states of quantum emitters, a canonical example being a single atom in a cavity (Fig. 1A). The circuit QED systems offer another variant of an elemental polariton: a single microwave photon in a resonator coupled to a two-level system (e.g., a superconducting qubit)[10,11] (Fig. 1C). Other actively studied realizations of polaritons involve solid-state emitters: in particular, color centers in diamond. Polaritonic phenomena have also been experimentally investigated in systems where light is coupled to single molecules[12], atomic clusters[13], or single atomic planes of van der Waals materials[2,14,15]. Despite these distinct physical implementations, polaritons in AMO, circuit QED and QM settings submit to a number of unifying principles rooted in the hybrid character of light-mater quasiparticles and even display common effects (Sections 3 and 4).

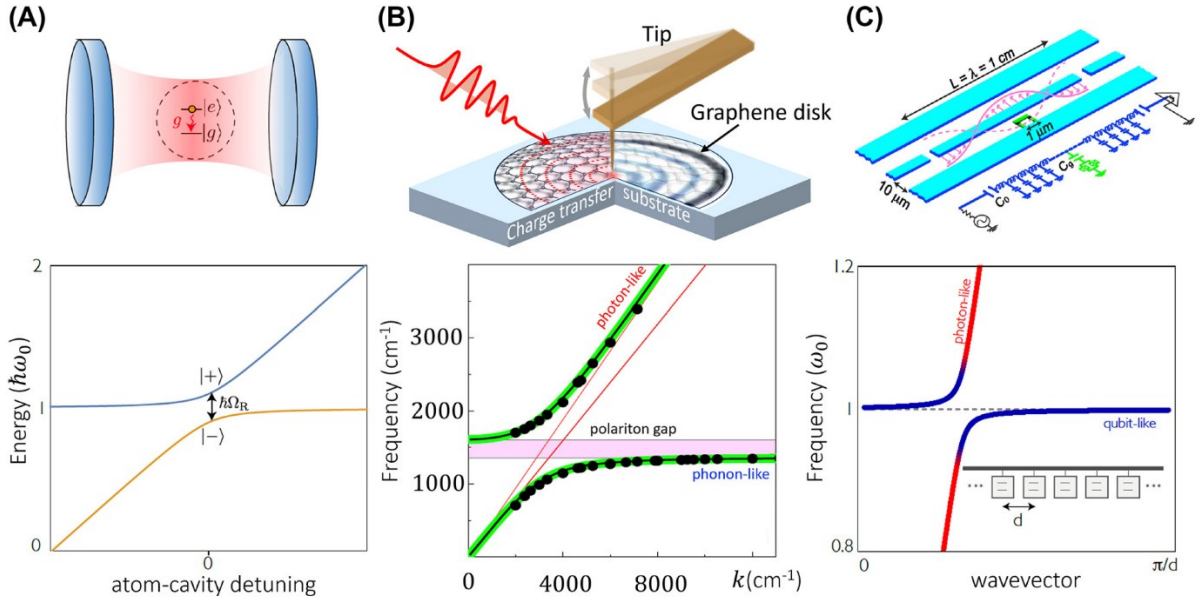

**Figure 1:** Polaritons formed by atoms (A), solids (B) and circuits (C). The top panels are schematic illustrations, and the bottom panels show the polariton dispersions (energy vs. detuning or energy vs. momentum/wavevector), revealing the hybrid light-matter character of polaritons. Panel A: Dressed states of a quantum emitter in a cavity. The hybridization between the atom and the cavity prompts the two distinct branches, with a vacuum Rabi splitting of $\hbar\Omega_R$. Panel B: Top: the schematic of a nano-optical experiment probing propagating plasmon polaritons using a scattering scanning near-field optical microscope (s-SNOM). Bottom: The coupling of a photon and a dipole-active resonance (plasmon, phonon, exciton, magnon, etc.) in a solid leads to the characteristic mode repulsion and energy partitioning into two branches of the frequency-momentum $(\omega,k)$ dispersion[16]. Panel C: A circuit QED system made of superconducting qubits embedded in microwave resonators. Top: cavities in circuit QED are formed by a transmission line hosting a superconducting qubit that acts as the "artificial atom." The qubit is placed at the middle of the resonator to couple to the strong electric fields at the antinode of the second mode. Adapted from Ref.[17]. Bottom: Polariton dispersion for an array of qubits coupled to a transmission line. Adapted from Ref.[18].

## 1B. Matter-like light, light-like matter and emergent properties of polaritonic quasiparticles

Polaritonic light-matter hybridization endows photons with matter-like properties, including nonlinear responses[19,20]. The acquired matter-like character effectively makes photons massive[21], enables optical processes that cannot occur in vacuum[22] and unlocks highly sought-after photon-photon interactions[11,20,23], setting the stage for many-body physics with light. Notably, the matter component of polaritons (quantified by the Hopfield coefficient $v$) can be tuned by scanning through the dispersion with varied frequency of the light field. The matter component can be altered by changing temperature[24,25]; by applying static electric[2,26], magnetic[27], or optical[28,29] excitation; by utilizing dielectric or metallic screening[30]; or by injecting electrical currents[31,32].

The photonic part of polaritons prompts long-range polariton propagation, entanglement[33,34], coherence and lasing[35,36]. Polaritonic rays in media generally abide by the rules of geometrical optics. However, the velocity and directionality of polariton travel in media can be strongly modified compared to free-space photon propagation. A striking example of unconventional polariton propagation is offered by conical ray trajectories and sub-diffractional focusing[37,38], lensing[39], as well as negative reflection[40], negative refraction[41,42] and polaritonic cloaking[43] in hyperbolic media. Photon chirality may induce helicity in exciton-polaritons in microcavities[44,45] or in propagating phonon polaritons[46]. The handedness of circular polarization enables control of the direction of plasmon polariton travel along the surface of patterned metallic nanostructures[47]. Optical anisotropy also introduces the photonic analog of spin-orbit coupling[48] (SOC), which may become imprinted onto the polaritons.

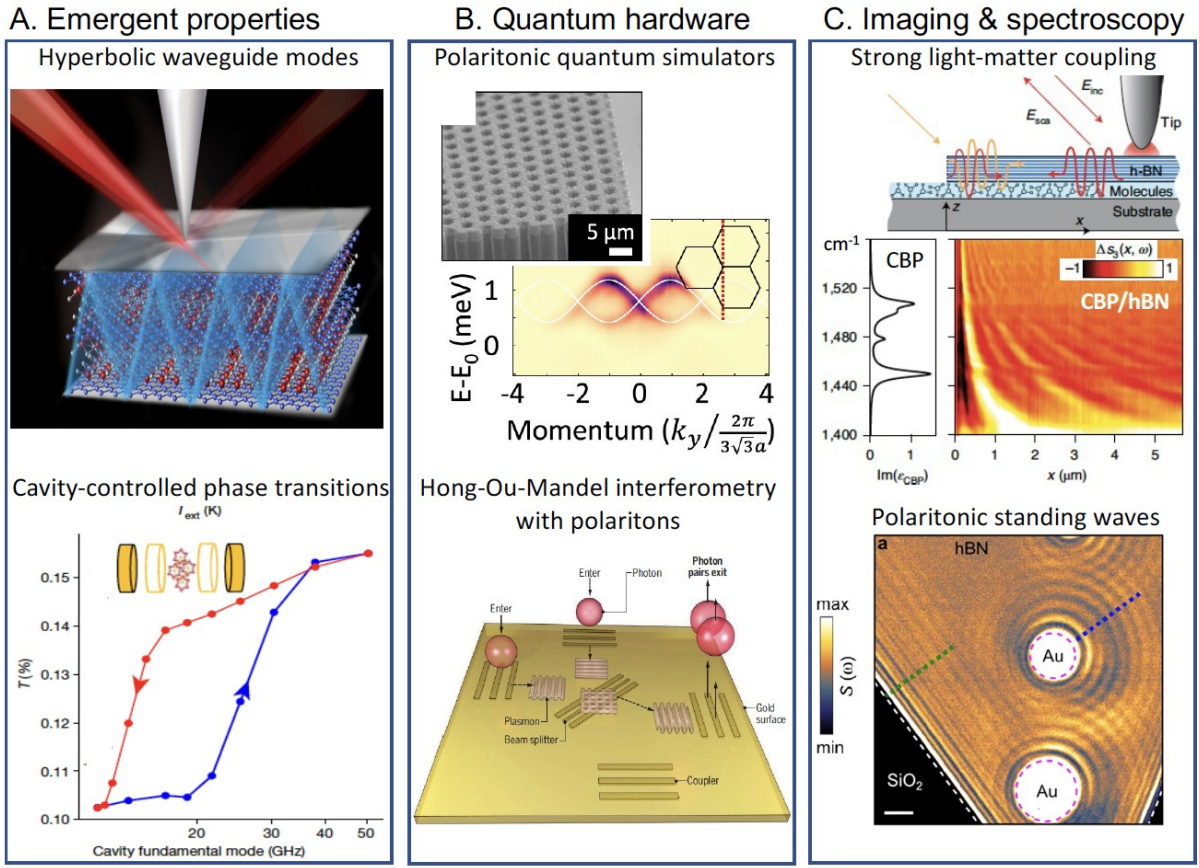

**Figure 2.** Emergent properties and functionalities in polaritonic systems. Panel A (top): Hyperbolic plasmon polaritons propagate through bulk crystals of the layered anisotropic metal ZrSiSe. Schematic adapted from Ref.[49]. Panel A (bottom): Reversible cavity control of the metal-to-insulator transition in 1T-$TaS_2$ integrated into the Fabri-Perot resonator (inset). The sample is kept at the fixed external temperature T = 150 K. The hysteresis as a function of the cavity fundamental mode is plotted as the evolution of the integrated low-frequency THz transmission (0.2 THz < ω < 1.5 THz). Adapted from Ref.[50]. Panel B (top): Exploring topological invariants in a polaritonic analog of graphene. Adapted from Ref.[51]. Scanning electron microscopy image of a honeycomb lattice of coupled micropillars giving rise to exciton-polaritons with Dirac-like dispersion. The main panel displays momentum-resolved emission spectra of a polaritonic graphene lattice. Panel B (bottom): Experiments with single photon polaritons show that quantum interference is preserved in the macroscopic polaritonic state involving ~$10^{10}$ electrons (Section 4). Quantum properties of plasmon polaritons were witnessed by implementing Hong-Ou-Mandel (HOM)

interferometry (Ref.[52]). Image adapted from Ref.[53]. Panel C (top): Strong coupling between propagating phonon polaritons and organic molecules. Experiment schematic reveals the concept of phonon-polariton interferometry of molecular vibrations. $E_{inc}$ and $E_{sca}$ denote the electric fields of the incident and tip-scattered radiation. The orange arrow indicates the illumination of the hexagonal boron nitride (hBN) edge hosting phonon polaritons. Molecular vibrations with 4,4′-bis(N-carbazolyl)-1,1′-biphenyl (CBP) molecules strongly couple to phonon polaritons in hBN, leaving characteristic resonances in the hyperspectral images (main panel). Adapted from Ref.[54]. Panel C (bottom): Nano-infrared image of polaritonic standing waves. Phonon polaritons in hBN are launched by Au antennas and detected with the help of a scattering nano-optical microscope (SNOM). Scale bar: 2000 nm. Adapted from Ref.[55].

*On polaritonic quasiparticles.* In physics, novel quasiparticles lead to novel macroscopic properties. For instance, in condensed matter systems, one witnesses unusual metallic transport in intermetallic compounds where s- and f-band electrons hybridize to form heavy fermion quasiparticles with giant effective masses $m_* \sim 1000\, m_e$, where $m_e$ is the free electron mass. Likewise, novel electronic and optical behaviors are observed in two-dimensional graphene or three-dimensional Weyl semimetals where quasiparticles are massless or nearly massless. Polaritonic quasiparticles also lead to a myriad of novel properties and responses, with several examples displayed in Fig. 2. For example, hyperbolic plasmon polaritons in anisotropic metals acquire the ability to propagate through the bulk of crystals in the form of waveguide modes (Fig.2A, top and Section 2B). Electron-photon states activated when materials are integrated in Fabry-Perot cavities alter the temperature of the insulator-to-metal transition (Fig.2A, bottom and Section 2C). Because interactions between polaritons can be controlled through either their light or their matter components, polaritonic systems promote programmable many-body effects that can be utilized in quantum simulators (Fig.2B, top and Section 3D). Polaritons created with entangled photon pairs preserve the key properties of their light constituents, including single- and two-photon interference[56] (Fig.2B, bottom and Section 4). Notably, interference of polaritonic quasiparticles enables novel means of inquiry into condensed matter physics rooted in the analysis of standing wave patterns (Fig. 2C and Section 2D). We will mention as a curiosity that polaritonic quasiparticles of either phonon or magnon origin are predicted to be sensitive to axion-photon coupling and thus may enable novel approaches in the search for dark matter[57,58].

**Box 1: Polariton family values and metrics**
*Vacuum Rabi splitting.* A quantitative measure of the light-matter interaction strength is the vacuum Rabi splitting $\Omega_R = \Omega_+ - \Omega_- = 2g = 2\mu E_{vac}/\hbar$ where µ is the transition dipole moment and $E_{vac}$ is the square root of the variance of the electromagnetic field in the vacuum state. Diverse polaritonic platforms allow one to achieve regimes of strong light-matter coupling[59–62] where coherent interactions overcome dissipation and decoherence. In the regime of strong light-matter coupling, the excitation is coherently exchanged between matter and field degrees of freedom. In cavity QED, strong coupling is achieved whenever g >κ, Γ$_0$, where *g* is the atom-photon coupling strength and Γ$_0$ and κ are the atom and cavity linewidths, respectively. The magnitude of κ dictates the cavity quality factor Q, which can exceed $10^{7}$[62]. Because light-matter coupling $g = \mu\sqrt{\frac{\omega_c}{2\epsilon_0 \hbar V}}$ scales as the inverse square root $V^{-1/2}$ of the polaritonic mode volume, the use of metallic nanostructures and plasmonic cavities[63] has emerged as a versatile approach for promoting hybrid states of light and matter (Box 3). Even if no photons are present in the polaritonic cavity, there is still an energy splitting known as vacuum Rabi splitting due to quantum fluctuations.

*Giant electrical dipoles promote strong coupling.* The coupling strength of light and matter can be enhanced in a solid-state system, revealing "colossal" dipole moments[62] relative to atomic platforms. The Rabi frequency scales as $\Omega_R \propto N^{1/2}$ where $N$ is the number of emitters within the mode volume $V$. Some of the strongest light-matter coupling strengths in solids $\Omega_R > 300$ meV have been reported for exciton-polaritons hosted by carbon nanotubes[64]. Even stronger Rabi splittings in excess of 500 meV are observed in molecular systems[65]. Atomic systems in their highly excited states (Rydberg states) also reveal the enhancement electric dipole moment μ that scales as the square of their principal quantum number ($n^2$). Rydberg physics relevant to polariton in solids (Section 4C).

*Energy scales.* Atomic, molecular, and solid-state systems offer diverse routes for building up matter polarization that collectively cover an extraordinarily broad range of the electromagnetic spectrum (Fig.3). Since photons of essentially any energy are widely available, polaritonic energy scales are dictated by the matter constituents. Even if the matter component can be approximated by a two-level system (atoms, qubits, excitons, Box 2), light-matter dressing adds an effective bandwidth to the polaritonic dispersion (lower panels in Fig. 1).

*Length scales.* A salient outcome of light-matter hybridization is that the polariton wavelength $\lambda_p = 2\pi/q_1$ is reduced compared to the wavelength $\lambda_0$ of light in free space (here $q=q_1+iq_2$ is the complex wavevector). In quantum materials, polaritonic confinement $\lambda_0/\lambda_p$ can be as high as $10^2$ to $10^3$. The electric fields are accordingly enhanced, as are the optical nonlinearities. Polaritonic wavelength in QMs can be directly read out from nano-optical images that visualize standing waves formed by propagating polaritons in space (Fig.2D) and in time (Fig.6 C-F).

*From strong to ultra-strong and deep strong light-matter coupling.* The cooperative enhancement of light-matter coupling, combined with large dipole moments in solids, is ideal for the implementation of strong light-matter coupling. In the weak coupling regime, radiative properties of atoms and solids are modified, with the Purcell effect being a notorious example. In the strong coupling regime, hybrid light-matter states are formed, provided the two constituents exchange energy faster than any dissipative processes. In the ultra-strong coupling regime, the coupling strength reaches a substantial fraction (typically defined as 20%) of the transition frequency, and the rotating wave approximation breaks down (Box 2). In the deep strong regime, the coupling strength exceeds the resonance energies, and the coherent coupling becomes the dominant energy scale in the system[7,61,62].

*Cooperativity, optical depth, and Purcell factor.* In cavity QED, the cooperativity C= $g^2/\kappa\Gamma_0$ quantifies the ratio between the emission into the cavity mode (at a rate of $g^2/\kappa$) and the decay into free space (at a rate of $\Gamma_0$). The cooperativity typically determines the performance of quantum information protocols implemented via light-matter interfaces, such as photonic gates, quantum memories[66,67], or quantum-enhanced metrology[68]. In waveguide QED, the optical depth (OD) $\Gamma_{1D}/\Gamma_0$ plays the same role, where $\Gamma_{1D}$ is the decay into the waveguide mode. A large optical depth signifies that an emitter coupled to a waveguide will mostly decay into the waveguide mode rather than to free space and also that a photon traveling through the waveguide will be almost perfectly reflected upon interaction with an emitter. For low optical depths (where reflection is low), a drive field will be exponentially attenuated as it propagates along the waveguide, with an attenuation coefficient that scales with the optical depth. Both the cooperativity and the optical depth are related to the Purcell factor $F_P$. For an emitter integrated in cavity with the mode volume V and the quality factor Q, the Purcell factor is defined as $F_p = \frac{4}{4\pi^2}\left(\frac{\lambda_0}{n}\right)\frac{Q}{V}$, where $\lambda_0$ is the wavelength in vacuum and *n* is the refractive index of medium inside the cavity. Large values of are observed for emitters in plasmonic nanocavities, which exhibit an increase in their spontaneous emission rates by a factor of $>10^3$ [36,69].

## SECTION 2: Polaritonic platforms

In this section we review the fundamentals of polariton physics and discuss common experimental implementations of polaritons. Despite vast dissimilarities among their matter constituents, polaritonic quasiparticles display many common traits.

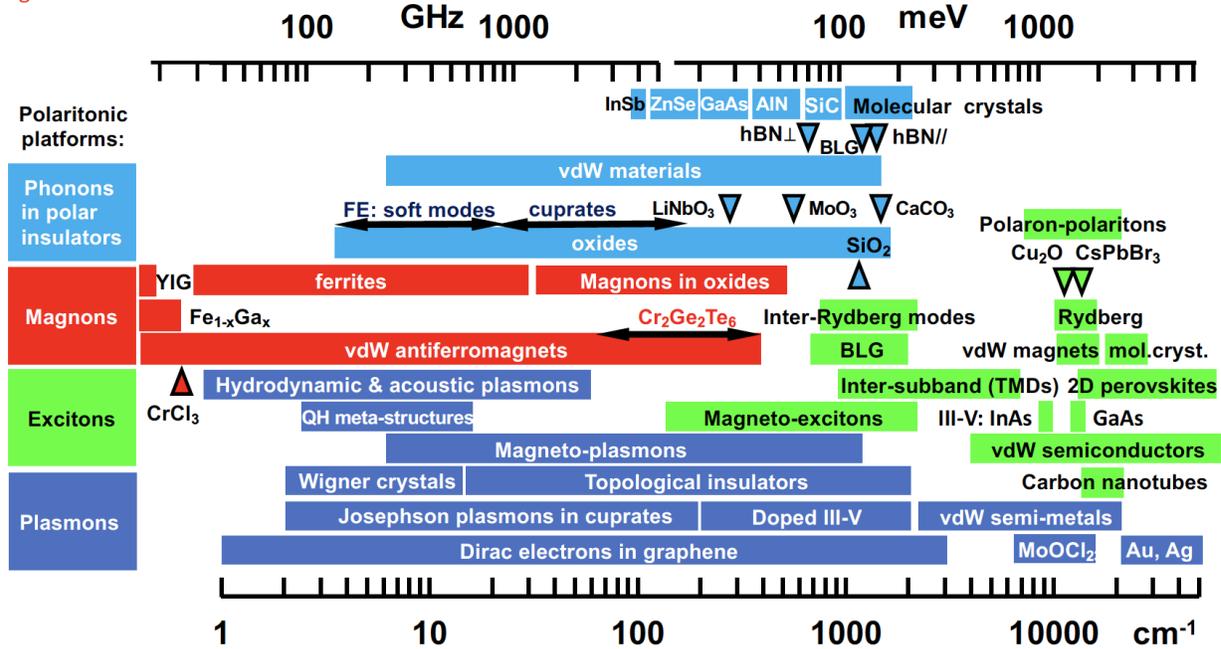

**Figure 3: Polariton taxonomy.** Polaritons occur in a wide assortment of physical settings. Characteristic frequencies of matter resonances span the range from GHz to Peta-Hz. Light-matter coupling physics is largely agnostic with regard to the microscopic origins of dipole-active matter resonances. Selected references for plasmon polaritons: Landau polaritons in GaAs/AlGaAs Hall bar integrated in a meta-structure [Ref.[70]]; Wigner crystals (theory) [Ref.[71]]; topological insulators [Refs.[72–79]]; Josephson plasmons [Refs.[80,81]]; vdW semi-metals [Refs.[49,82]]; graphene plasmons [Refs.[26,30,83–85]], surface plasmon polaritons in elemental metals [Refs.[8,86]]; hydrodynamic and acoustic plasmons in quantum materials [Refs.[87–90]]; alkalai metals [Ref.[79]]; hyperbolic plasmons in MoOCl$_3$ [[9192]]; QH refers to the quantum Hall effect. Selected references for exciton-polaritons [Ref.[93]]: perovskites [Refs.[94–96]], carbon nanotubes [Ref.[64]]; vdW semiconductors [Refs.[14,25,97–99]]; vdW magnets [Refs.[100,101]]; molecular crystals [Refs.[102,103]]; Rydberg [Refs.[95,104]]; inter-Rydberg modes [Ref.[28]]; III-V semiconductors [Refs.[105–107]]; CrSBr [[101]]; bilayer graphene [Ref.[108]]; polaron-polaritons [Ref.[109]]. Selected references for magnon polaritons: transition metal oxides [Refs.[110,111]]; vdW antiferromagnets [Ref.[112]]; ferrites [Refs.[113,114]]; yttrium iron garnet (YIG) [Refs.[115,116]]; Fe$_{1-x}$Ga$_x$ [Ref.[117]]. Selected references for phonon polaritons: van der Waals materials [Refs.[26,118]]; various oxides [Ref.[119]]; SiO$_2$ [Ref.[120]]; MoO$_3$ [Refs.[121,122]]; SrTiO3 [Ref.[123]]; LiNbO$_3$ [Ref.[124]]; Y$_2$SiO$_5$ [Ref.[125]]; LiV$_2$O$_5$ [Ref.[126]]; β-Ga$_2$O$_3$ [Ref.[127]]; ATeMoO6 (A = Mg, Zn, Mn, Co, etc.) [Ref.[128]]; CdWO$_4$ [Ref.[129]]; EuFeO$_3$ [Ref.[130]]; CaCO$_3$ [Ref.[131]]; BaTiO$_3$ [Ref.[132]]; BiFeO$_3$ [Ref.[133]]; ferroelectric (FE) soft modes [Ref.[134]]; SiC [Refs.[119,135]]; bilayer graphene (BLG) [Ref.[136]]; molecular crystals [Refs.[137,138]].

## 2A. Polaritons in quantum materials

Polaritons are omnipresent in quantum materials[2]. Dipole active polarization channels, including NV-center spins, superfluids, magnons, plasmons, excitons, and phonons, among others, all give rise to prominent resonances that can hybridize with light (Fig.3).

*On higher-order polaritons.* In quantum materials, multiple excitations with overlapping frequencies, simultaneously present in the same material or in proximal layers in heterostructures, support higher-order polaritons, including plasmon-phonon and plasmon-exciton polaritons[2,139–141]. A number of higher-order polaritons have been discovered in 2D van der Waals (vdW) heterostructures[30,142] where ultrathin layers with dissimilar dielectric response (insulators, semiconductors, and metals) are assembled together (Fig. 4A-C). The discovery of Landau phonon polaritons (Fig. 4B) in graphene/hBN structures[143] was enabled by the development of infrared nano-imaging in high magnetic fields, which also revealed nano-optical coupling to magneto-excitons[144] (see also Section 3F on magneto-polaritons). Fig. 1B (lower panel) displays the result of the hybridization of infrared photons with a weakly dispersing matter resonance (phonon). The dispersion of higher-order polariton modes is often more complex (Fig. 4A-C): it can have multiple branches, multiple avoided crossings, and non-monotonic momentum dependence (negative group velocity). The flat part of the dispersion implies that high-momentum polaritons in quantum materials effectively behave as heavy photons with dramatically reduced group velocities. The details of the dispersion and therefore the velocity of the polaritons in quantum materials[30,142] can be controlled by, e.g., activating the dielectric screening of the polaritonic medium using proximal layers with high dielectric function, including highly conducting metallic layers.

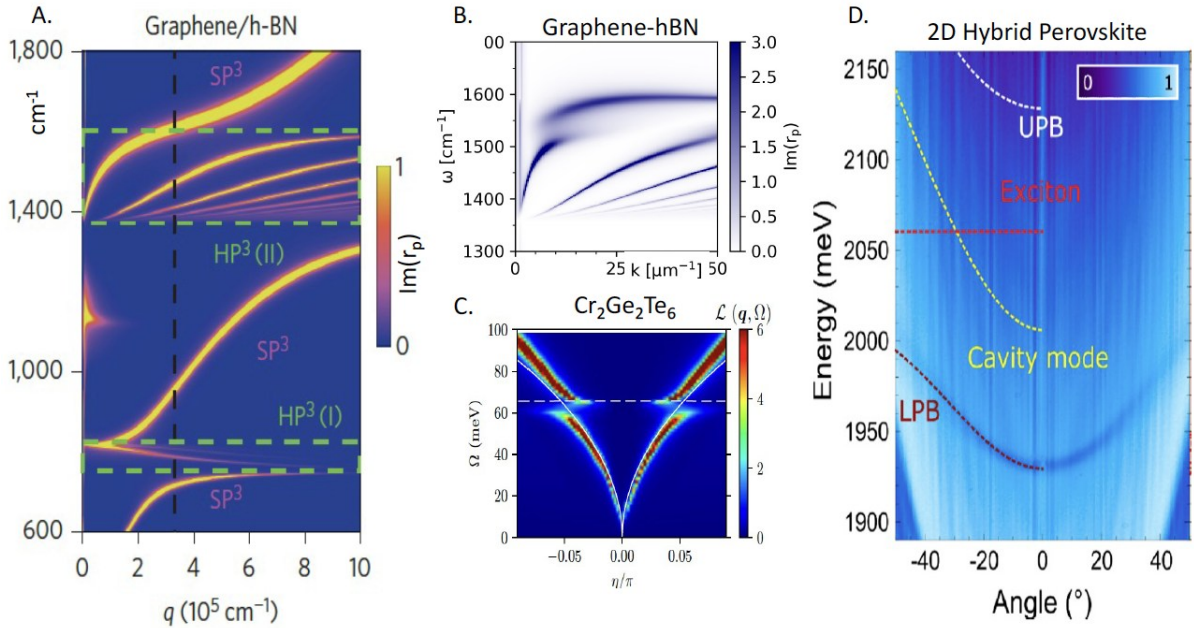

Figure 4. Energy-momentum dispersion relationships. Panel A: Calculated dispersion of hybrid polaritonic modes in a graphene-hBN heterostructure. The dispersion is visualized using a false-color map of the imaginary part of the reflection coefficient $r_p$. The black dashed line is a rough estimate of the momentum at which the tip-sample coupling is strongest. Green dashed rectangles surround the regions of the hyperbolic response of hBN. The false-color map reveals the dispersion of the hyperbolic plasmon-phonon polaritons (HP3) and the surface plasmon–phonon polaritons (SP3). Weak resonances around $\omega = 1,130$ cm$^{-1}$ originate from the SiO$_2$ substrate. Adapted from Ref.[145]. Panel B: Hybridization of hBN phonon polaritons with graphene Landau polaritons (LPPs). Calculated LPP dispersion at the magnetic field of 3.35 T reveals the field-induced gap in the dispersion. The false color represents the imaginary part of the complex reflectance $r_p$. Adapted from Ref.[143]. Panel C: The calculated electron energy-loss function $L(q,\Omega)$ (color scale), revealing magnon-plasmon dispersion for a monolayer of the doped antiferromagnetic semiconductor $Cr_2Ge_2Te_6$. Plasmon-magnon branches originate from electron-magnon interactions. Adapted from Ref.[146] Panel D: Exciton-polariton dispersion of a 2D hybrid-perovskite semiconductor integrated in the microcavity. The dispersion data are presented in the form of an angle-resolved reflectivity map. The dashed white and dark-red curves are the upper polaritonic branch (UPB) and lower polaritonic branch (LPB), respectively, obtained by using a two-coupled oscillator model to fit the experimental data. The dashed red line and dashed yellow parabola represent the exciton energy and the dispersion of the bare cavity mode, respectively. Adapted from Ref.[96]

*Cavity polaritons.* Semiconductors integrated in cavities are an extensively studied form of polaritonic quantum matter. Here, the term "cavity" colloquially describes a collection of modern micro- and nano-optical structures (Box 3) that play a role similar to that of their Fabry-Perot ancestors (Box 2). The principal function of cavities in the weak coupling regime is to modify the photonic density of states experienced by quantum materials, as attested by the Purcell effect and the excitonic Lamb shift[62]. Strong coupling between cavities and excitons in semiconductors gives rise to (micro)cavity exciton-polaritons as exemplified in Fig.4D. The effective mass of polaritonic quasiparticles is small compared to free electron mass (down to $\sim 10^{-4}$ $m_e$) but remains finite, at variance with massless photons. Cavity exciton-polaritons interact with each other; these interactions are rooted in their excitonic constituent[23]. Microcavity exciton-polaritons reveal Bose-Einstein condensation (BEC), superfluidity, lasing, quantum vortices and many other effects seen in cold-atom BECs[45,147]. Apart from BEC, strong light-matter coupling regimes have been experimentally achieved between "cavity photons" and various material resonances, including, but not limited to: Landau level transitions in 2D electron gases[27], interband transitions[148], vibrational resonances[54] and intrinsic Josephson resonances in layered high-$T_c$ superconductors[149].

*On bright and dark polaritons.* The notion of dark polaritons is introduced in the vast polaritonic literature in multiple contexts. First, both organic and inorganic material host states that only weakly interact with light. In fact, the physics of organic/molecular systems integrated in microcavities is governed by a continuum of dark states. The Tavis-Cummings model (Box 2) dictates that a system of $N$ molecules and one photon has $N+1$ degrees of freedom; $N-1$ of these modes are dark, in addition to the two bright polaritons[150]. In bulk organic microcavities, the dark states dominate over the bright polaritons on a scale of $\sim 10^5$. Dark states play a prominent role in a wide array of properties[138,151,152], including the functioning of common organic light-emitting devices[153]. A different commonly encountered use of the term dark states refers to dark excitonic states that are common in inorganic semiconductors. For example, in transition metal dichalcogenides, strong spin-orbit coupling (SOC) leads to spin- and energy-splitting in the conduction band. The electrons with anti-parallel spins in the two respective states of the split conduction band give rise to bright and dark excitons[154]. The connection to polaritons is that dark exciton can be experimentally accessed via near-field coupling facilitated by surface plasmon polaritons[155] or by nano-plasmonic cavities[156,157]. Next, dark-state polaritons are introduced in the context of mixtures of photonic and Raman-like matter modes[158]. The dark-state polaritons give rise to ultra-slow light and electromagnetically induced transparency (Section III.A). Finally, a variety of polaritonic strong coupling scenarios result in dispersion branches situated outside the light cone (Fig.1B).These latter highly confined, high-momentum polaritons are "dark" because of the momentum mismatch. High-momenta dark polaritons but can be investigated using a variety of nano-optical imaging methods (Fig.2C), as we discuss in Section 2D.

*On cavity and Floquet dressings.* Cavity dressing can occur concurrently with laser-induced Floquet dressing of the electronic bands in a semiconductor[159,160]. Despite the distinction between cavity dressing and laser/Floquet physics, the end result of either scenario is a dressed polaritonic state. Notably, laser and cavity dressing can work in concert to create a new quasiparticle, the phonoriton[139], consisting of three components: photons, excitons and phonons (Box 2). Quantum materials integrated in laser-driven cavities reveals pronounced changes in the polaritonic dispersion, manifested in strong nonlinearities[161]. Furthermore, capitalizing on cavity engineering of the vacuum modes, it has become possible to achieve orders-of-magnitude enhancement of the effective Floquet field, enabling Floquet effects at an extremely low fluence of 450 photons/$\mu m^2$[162]. At higher fluences, the cavity-enhanced Floquet effects led to 50 meV spin and valley splitting of $WSe_2$ excitons, corresponding to an enormous time-reversal-breaking, non-Maxwellian magnetic field exceeding 200 T.

## 2B. Polaritons in van der Waals materials

Van der Waals materials host a broad variety of polaritons and polaritonic effects that have already been reviewed[26,84,118]. Here we outline several results that, for the most part, came into the limelight after the above comprehensive reviews had been published; these include studies of self-hybridized cavities/polaritons, anisotropic and hyperbolic waveguiding, and interface and moiré engineering.

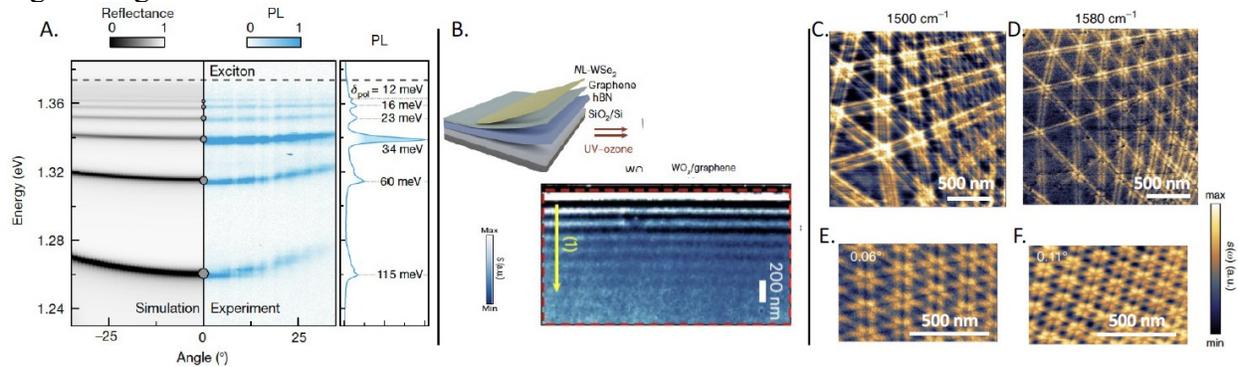

Figure 5. Polaritons in van der Waals material and heterostructures. Panel A: Self-hybridized exciton-polaritons in a slab of CrSBr single crystals. The right panel shows the angle-resolved and angle-integrated photoluminescence (PL) of a 580-nm-thick crystal recorded at T = 1.6 K. The left panel shows the simulated reflectance map for conditions matching the experiment. Adapted from Ref.[163]. Panel B: Charge transfer plasmons in a graphene-$WO_x$ heterostructure. Top: charge-neutral graphene is covered with $WSe_2$. Subsequent oxidization transforms the topmost monolayer of $WSe_2$ into the high work-function material $WO_x$. Bottom: nano-infrared image of the scattering amplitude $S(r,\omega)$ of $WO_x$/graphene obtained at T = 300 K and $\omega$ = 980 $cm^{-1}$. Scale bars: 200 nm. Adapted from Ref.[164]. Panels C-F: Nano-IR imaging of moiré patterns in twisted bilayer graphene encapsulated with hBN. Data plotted in the form of the normalized scattering amplitude (Panels C and D) for structures with a twist angle of ~$0.05^0$. Scattering amplitude data collected at 1560 $cm^{-1}$ for structures with twist angles of $0.06^0$ and $0.11^0$ are plotted in Panels E and F, respectively. Adapted from Ref.[165].

*On self-hybridized polaritons.* Crystals of vdW quantum materials can serve as their own cavities or resonators, a feat enabled by their high dielectric permittivity (Box 3). The optical modes of these slab cavities can hybridize with excitons in the same material, leading, for example, to self-hybridized polaritons[163,166,167]. Self-hybridized exciton-polaritons in a single crystal of CrSBr are displayed in Fig. 5A. The energy-momentum dispersion of the photonic waveguide modes can be readily controlled by selecting the crystal thickness. These thickness-dependent dispersions of vdW waveguides were utilized to satisfy momentum-matching conditions for second harmonic generation and spontaneous down conversion[168–171].

*On hyperbolic insulators, semiconductors and metals.* Another common property of vdW crystals is their hyperbolicity. In hyperbolic materials permittivity have opposite signs along different crystal axes[172–175]. Hyperbolicity gives rise to high-momentum waveguide modes. This effect was first observed in hBN, where hyperbolic polaritons are of phonon origin[37,176]. Phonon polaritons have been identified in a broad variety of anisotropic insulators[177]. The library of hyperbolic materials is rapidly expanding[175]: recent additions to the family of hyperbolic systems include hyperbolic exciton-polaritons in the vdW semiconductor CrSBr[101] as well as hyperbolic plasmon polaritons in the anisotropic metals ZrSiSe[49] and $MoOCl_2$[91,92]. Sternbach et al. were able to initiate a hyperbolic response of exciton-polariton origin in the common crystal $WSe_2$ by photo-excitation with an external pump laser[178]. Fu et al. utilized optical pumping of black phosphorous to initiate a transition to a hyperbolic response of plasmon polaritons[179]. Finally, He et al. succeeded in tuning hyperbolic phonon polaritons in hBN/InAs and hBN/CdO structures by ultra-fast photoexcitation of the substrates' semiconductors[180].

*On interface engineering: polaritonic band structures.* Van der Waal heterostructures offer versatile means to control polaritonic responses by judicious engineering of interfaces. Lundeberg et al. succeeded in tuning the velocity of plasmon polaritons in graphene by varying the spacer thickness between the graphene microcrystal and the Au substrate. With the spacer as thin as 5.5 nm, it became possible to lower the group velocity of the polaritons nearly down to the Fermi velocity, thus gaining access to non-local electrodynamics of graphene[30]. Patterned substrates can create a periodic potential for propagating polaritons. This approach has been extensively utilized to impose complex "band structures" for polaritonic waves[140,181–186]. Periodic structures created by direct patterning of vdW materials supporting polaritons likewise provide a means for controlling nano-optical effects in these systems. For example, Zhang et al. experimentally demonstrated extremely asymmetric and unidirectional phonon polaritons by directly patterning the high-symmetry orthorhombic van der Waals crystal $\alpha$-$MoO_3$[187].

*On interface engineering: charge transfer polaritons.* Van der Waals materials reveal a broad range of work functions: from some of the lowest ones (~2 eV) in $ZrO_x$ to some of the highest ones (~6.5 eV) in $RuCl_3$. An interface of dissimilar materials with distinct work functions may acquire an excessive charge and develop substantial conductivity. Rizzo et al. harnessed charge transfer at the interface of graphene with $RuCl_3$ to initiate plasmonic response in otherwise undoped and ungated structures[188]. The analysis of the polaritonic dispersion revealed a Fermi energy $E_F$=0.6 eV and a plasmonic quality factor comparable to that of the highest-quality encapsulated devices. The charge transfer process creates the abrupt variations in the Fermi energy in graphene that are often required for nano-photonic structures[189,190]. Kim et al. demonstrated ambipolar charge transfer in graphene/$ZrO_x$ and graphene/$WO_x$ structures (Fig.4B). Rizzo et al observed unidirectional plasmons prompted by charge transfer at the interface of graphene with anisotropic magnetic semiconductor CrSBr[191]. Finally, Sternbach et al. were able to harness charge transfer at the $WSe_2$/$RuCl_3$ interface to quench photo luminescence and prompt plasmonic response of $WSe_2$ in THz frequencies.

*On polaritonic moiré superlattices.* Van der Waals materials display a broad variety of moiré superlattices. By varying the twist angle $\theta$ between adjacent layers one can control the period of superlattices. Moire superlattices are also nearly universally applicable for programming electronic, magnetic, excitonic, photonic and other phenomena[192]. Two identical lattices with a relative twist (of angle $\theta$) or two similar lattices with a small difference in their lattice constants can be superimposed to produce a moiré pattern. The wavelengths of the moiré pattern can exceed many unit cells of the constituent materials. Within the moiré mini-Brillouin zone, the folded bands lead to a multitude of crossings, which are subsequently split via interlayer hybridization, resulting in band flattening and the quenching of kinetic energy. Likewise, moiré periodicity creates the band structure for propagating polaritons in a fashion fully analogous to artificial photonic crystals[193]. This leads to rich interference patterns in structures with ultra-small twist angle [165,193,194] exemplified in Fig.5C-F. Moire bilayers with $\theta \sim 1.1$-$1.7^0$: the so-called magic angle, display plasmonic response attributable to interband transition between the mini bands of the moire superlattice[195]. Moiré superlattices have been proposed as a platform for ordered excitonic arrays with excitonic/polaritonic interactions[196–199] and nonlinearities[99] controlled by the angle $\theta$. Furthermore, the moiré platform is well suited to create gate-tunable Feshbach resonances, thus providing the basis for tunable interactions for bosonic particles (excitons and polaritons) as well as for fermionic electron and hole states[197]. Twisted TMD materials, including $WSe_2$, reveal ferroelectric polarization associated with the moiré domains. Zhang et al. integrated graphene monolayers integrated in ferroelectric structures based on minimally twisted $WSe_2$ and hBN; they harnessed plasmons in graphene to visualize alternating polarization associated with neighboring moiré domains[200,201].

## 2C. Cavity materials engineering and vacuum state polaritons

Matter can hybridize with the vacuum state of a cavity. Thus, no external photon pumping is required to promote coupling of the cavity mode with vibrational and electronic degrees of freedom, giving rise to vacuum field polaritonic states. This is not surprising, as vacuum fluctuations are the origin of well-known phenomena such as spontaneous emission, van der Waals interactions, the Casimir effect, and many others. As in earlier examples in Sections 1-2 and Box 3, the term "cavity" here refers to a plethora of implementations of (nano)structured electromagnetic environments surrounding atomic and solid-state emitters, materials or qubits. The significance of vacuum states in polaritonic effects can be appreciated by considering the expression relating the Rabi frequency experienced by a medium with dipole moment $d$ embedded in a cavity to the number of photons $n_{ph}$ in the cavity[52]:

$$\hbar\Omega_R = 2d\sqrt{\frac{\hbar\omega}{2\varepsilon_0 v}} \times \sqrt{n_{ph}+1} \quad (1)$$

where $\varepsilon_0$ is the permittivity of the vacuum and $v$ is the mode volume. The second term in Eq.1 implies that the Rabi splitting is finite even in the absence of photons in the cavity.

*On cavity quantum materials.* Some of the most exciting properties of quantum materials arise from strong interactions among electrons, spins, and crystal lattices[202]. A question in the vanguard of current interest and debate is whether the strong polaritonic coupling of cavity modes with material resonances can also prompt new quantum phases and novel states of matter. If so, placing a quantum material in a cavity and/or in a photonic structure (Box 3), and thereby enhancing the light-matter coupling, may potentially lead to new material properties even in the absence of any external optical field. This line of research is rooted in the successes of cavity QED: atoms, molecules and superconducting qubits (commonly referred to as artificial atoms) do reveal strong coupling to the cavity vacuum modes, as witnessed by the Rabi splitting. The possibility of achieving cavity control of ground-state properties is yet to be fully understood at the theoretical level. It has been argued that a no-go theorem excludes quantum phase transition in materials integrated in a cavity[203–206]. All these ideas are subject to intense discussion and await resolution by direct experiments. Section 5 outlines some of the pressing open questions in cavity-altered materials.

*On cavity materials engineering: platforms, theory and modeling.* Theoretical predictions for cavity-QED control of quantum materials are enticing[45,62,207–214]. Cavity enhancement of the superconductivity[207,209,215–217], ferroelectricity[210,218], topological properties[219,220], and cavity-induced non-Fermi-liquid behavior of vdW materials[221] have been theoretically proposed. Strong light-matter coupling of molecules and quantum materials to cavity modes has been experimentally demonstrated[62]. We stress that plasmonic cavities and various nano-photonic structures, including the split-ring resonators (SRRs) displayed in Box 3, often provide some of the strongest coupling strengths[64,222–225]. Cavity electrodynamics can be implemented in multilayer structures of vdW materials by harnessing intrinsic resonance properties of constituent layers[81,226]. Advanced nano-fabrication methods applied to vdW materials have enabled the construction of cavity structures with enticing figures of merit[227,228]. It has been theorized that a coupling of a solid to a quantized cavity mode may renormalize the band-structure [229,230] into electron-polariton bands[231]. Furthermore, polaritonic chemistry offers remarkable experimental examples of how chemical reactions can be altered in molecular systems coupled to vacuum-state polaritons in cavities[138,232–237].

*On cavity materials engineering: first data.* Experimental progress in cavity materials engineering is also encouraging[4,7,232]. For example, GaAs quantum Hall bars were integrated with SRRs[238]. The strongly enhanced vacuum field fluctuations in SRRs altered transport properties in the quantum Hall state, leading to the breakdown of the topological protection. This work proposes that long-range hopping mediated by SRR cavities leads to finite resistivity that is ultimately induced by vacuum fluctuations in the correlated electron-photon platform. Enker et al. demonstrated that fractional quantum Hall gaps can be enhanced due to the vacuum field in a nearby hovering split-ring resonator (submitted). Jarc et al. reported on the reversible cavity control of the thermally tuned metal-to-insulator transition in a correlated solid-state material (Fig2A). The authors embedded the charge density wave material 1T-$TaS_2$ into a tunable terahertz cavity and were able to switch between conductive and insulating behaviors[50].

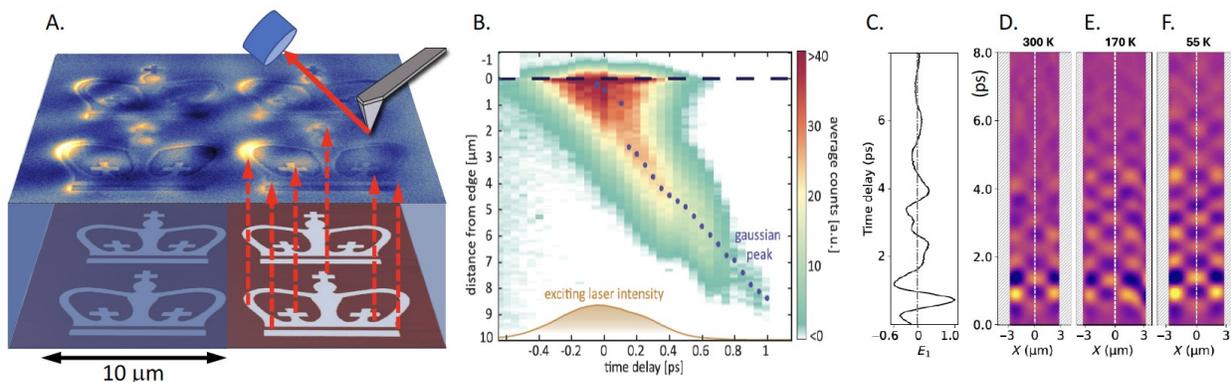

**Figure 6.** Imaging of propagating polaritonic wavepackets in space and time. Panel A: Canalization imaging with phonon polaritons in hBN. A lithographically defined pattern on a Si/$SiO^{17}$ substrate (the logo of Columbia University in the City of New York) is canalized by phonon polaritons across the hBN crystal. The image on the top surface is obtained with an s-SNOM operating at a wavelength of ~ 7 μm. Unpublished data by S.Sunku and D.N.Basov. Panel B: Map presenting the measured phonon polariton wavepacket in hBN as a function of time and distance from the edge of the crystal. The blue dots display the Gaussian wavepacket peak and gold represents the laser excitation intensity. Here, the time delay $t_d = 0$ is related to the peak of the excitation intensity. This experiment uses a 55-nm-thick hBN flake, excited by a 6470-nm

laser with a bandwidth of 175 nm. Adapted from Ref.[239]. Panels C-F: plasmonic worldline metrology applied to monolayer graphene. Panel C: THz time-domain signal collected with the tip located at the center of the graphene ribbon. Second-order spatial derivatives of spacetime maps taken for a 6-mm-wide graphene ribbon at 300, 170, and 55 K, all at gate voltage Vg = 15 V (Panels D-F). The checkerboard pattern is produced by the space-time interference of traveling plasmonic wavepackets. Adapted from Ref.[240].

2D. Propagation of high-momentum polaritons in space and time

The relative weights of the matter and photon excitations in a polaritonic state can be tuned such that the light dressing is negligible, with the high-momentum polariton becoming mostly matter-like or "dark," as we discuss in Section 2A. Even though the photonic component of the high-momentum polariton is small, it imbues matter with light-like properties. These dark polaritons propagate in the form of waves with a slow group velocity[2,62,241]. In the extreme case realized in the regime of electromagnetically induced transparency (Section 3A), polaritonic dressing can bring light to a standstill. In extended systems such as 2D materials, part of the dispersion relation settles to the right of the light cone (Fig. 1B,C), resulting in momentum mismatch with photons in free space or in the transmission line.

*Nano-optical near-field imaging.* Modern optical nano-imaging methods allow one to "see" predominantly matter-like polaritons in quantum materials by overcoming the momentum mismatch between free-space photons and high-momentum polaritons[242]. Plasmon polaritons in 2D conductors, which reveal a characteristic $\omega \propto \sqrt{q}$ dispersion positioned entirely outside the light cone, are a case in point. Optical antennas and nanostructures (Box 3) enable efficient coupling of light to such high-momentum modes, which remain dormant in conventional optical experiments. The propagation of plasmon polaritons in real space has been visualized in graphene and other van der Waals materials or structures (Fig.1B, Fig.2C). Polaritons display familiar optical phenomena – reflection, interference, and refraction[41,42,118,243] – but at length scales far below the diffraction limit. For example, when a plasmon polariton is confronted by an abrupt change in the dielectric function on its path, it becomes partially or completely reflected, generating rich interference patterns (Fig. 5C-F). Quasiparticle interference (QPI) is a common concept utilized in the scanning tunneling microscopy of solids[244–246]; the use of polaritonic interference and standing waves is a close counterpart of QPI in tunneling microscopy[24,83,247]. The lineforms of polaritonic standing waves encode information about the frequency dependence and momentum dependence of the response functions of media that support polaritons. Thus, polaritonic images allow one to inquire in a novel way into electronic, excitonic, magnetic, superconducting, and other interesting phenomena in quantum materials.

*On polaritonic waveguiding and hyperbolic modes.* Anisotropic quantum materials support polaritonic waveguide modes (Fig.2A) traveling through the bulk of single crystals as sub-diffractional polaritonic "rays"[37,173,248]. Nano-optical methods are capable of visualizing these so-called hyperbolic polaritons as they reach the crystal surface, thus allowing one to reconstruct polaritonic trajectories inside the host media. Hyperbolic media give rise to the effect of polariton canalization (Fig.6A): intrinsic collimation of energy flow along a single crystalline axis[249]. Polariton canalization enables the guiding of light in anisotropic media and holds promise for applications in imaging, lithography and directional thermal management, among others[37,250–254]. Canalization of polaritons can be implemented for in-plane transport of polaritonic wavepackets[255]. Moreover, twisted asymmetric stacks of hyperbolic polaritonic media support unidirectional ray-like propagation of phonon polarions[256].

*On space-time tracking of traveling polaritons: near field*. Several research teams combined nano-imaging methods with ultra-fast lasers, a combination that has enabled the tracking of polaritonic wavepackets not only in space but also in time, at sub-picosecond time scales[239,240,257–259]. Kurman et al. augmented an ultrafast transmission electron microscope with synchronized femtosecond mid-infrared lasers[239]. This latter experimental approach allowed the authors to execute spatiotemporal measurements of 2D wavepackets formed by phonon polaritons in hBN (Fig.6B). Yoxall et al. proposed time-domain interferometry also harnessing fs mid-IR lasers and were able to document negative phase velocity associated with the low-frequency phonon polariton branch in hBN[257]. Zhang et al. successfully captured movies of distinctive pulse spatiotemporal dynamics, including curved ultraslow energy flow trajectories, anisotropic dissipation, and dynamical misalignment between phase and group velocities, also in hyperbolic crystals of hBN[258]. Xu et al. reported on the space-time mapping of surface plasmon polaritons in the THz regime. The authors introduced the notion of polaritonic space-time worldlines directly accessible through THz nano-imaging of plasmon polariton wavepackets (Fig.6C-F). The analysis of worldlines yields plasmonic group velocities and lifetimes. This analysis uncovered novel aspects of electronic interaction in the Dirac electron fluid of graphene[240].

*On space-time tracking of traveling polaritons: far field*. Near-field space-time imaging is complemented by a growing arsenal of far-field techniques capable of tracking polariton propagation both within and outside the light cone[260–265]. An important realization stemming from spatiotemporal measurements of polariton propagation is that highly matter-like polaritons exhibit group velocities that are slower than predicted from the dispersion obtained in linear spectroscopies, and may even decohere on much shorter timescales than their lifetimes, due to interactions with lattice phonons or polaritonic dark states. Recent realistic theoretical treatments of polariton propagation[266–268] concur with these measurements and provide insight on the role of inevitable disorder in polariton dynamics.

*On Zitterbewegung or trembling motion of polaritons*. In quantum mechanics, the particle in motion induces the precession of the spin, which in turn affects the particle velocity, leading to a wiggling or trembling trajectory termed Zitterbewegung and first discussed by Schrödinger. This quantum relativistic effect entails the oscillation of the center of mass of a wavepacket in the direction perpendicular to its propagation. It was theoretically proposed that the Zitterbewegung of exciton-polaritons may originate from the optical constituent of the polariton state[269]. The splitting in the transverse electric and transverse magnetic cavity modes (TE–TM splitting) is at the origin of this effect. Experimental signatures of Zitterbewegung were detected via direct monitoring of exciton-polariton trajectories[270–272].

**Box 2: How do polaritons come to be?**

In this Box we briefly introduce some of the key models of light-matter interactions underlying the entire edifice of polaritonic physics.

*An elemental polariton* is formed by an atom hybridized with the optical mode of a high-$Q$ cavity in the strong coupling regime, where the coherent interactions between the atom and the cavity mode overcome the dissipation arising from photon leakage and spontaneous emission into other optical modes[273]. In this scenario (Fig. 1A), the atom can no longer be understood as an isolated entity and instead is said to be "dressed" by the electromagnetic modes of the cavity.

*Quantum electrodynamics* (QED) provides a general framework for treating electromagnetic and matter degrees of freedom on an equal quantized footing. In many cases, the full QED theory of light-matter interaction can be approximated by a simpler non-relativistic approach. Namely, for a system of $N_e$ electrons and $N_n$ nuclei with nuclear charge $Z_i$ and mass $M_i$, the corresponding non-relativistic Pauli-Fierz Hamiltonian[274] for M photon modes (of frequency $\omega_k$ and light-matter coupling $\lambda_k$) in the long-wavelength (dipole) approximation can be written as[275]:

$$\hat{H}_{PF} = -\sum_i^{N_e} \frac{\hbar^2}{2m_e}\nabla_i^2 + \frac{e^2}{8\pi\epsilon_0}\sum_{i,j=}^{N_e}\frac{1}{|r_i - r_j|} - \sum_i^{N_n}\frac{\hbar^2}{2M_i}\nabla_i^2 + \frac{e^2}{8\pi\epsilon_0}\sum_{i,j}^{N_n}\frac{Z_iZ_j}{|R_i - R_j|}$$
$$+ \frac{1}{2}\sum_k^M \left[\hat{p}_k^2 + \omega_k^2\left(\hat{q}_k - \frac{\lambda_k}{\omega_k}\hat{R}\right)^2\right] - \frac{e^2}{4\pi\epsilon_0}\sum_{i,j}^{N_e,N_n}\frac{Z_j}{|r_i - R_j|} \quad (1)$$

where the light-matter interaction is absorbed into an effective photonic-like component ($\hat{q}_k - \frac{\lambda_k}{\omega_k}\hat{R}$) linked to the transverse electric field of the photon mode $k$ and the canonical operator $\hat{q}_k$ is related to the corresponding displacement field. So far, only a few theoretical approaches, such as quantum electrodynamic density-functional theory (QEDFT)[276] and coupled-cluster theory[277], present feasible solutions for realistic systems. A common approximation is to omit and reabsorb the self-polarization term $\sim(\lambda R)^2$ into an adjusted physical particle mass.

*Two-level systems.* By expressing the dipole operator in a restricted two-level basis (with energy gap $\hbar\omega_{exc}$) and assuming a single cavity mode of frequency $\omega_c$, Eq. (1) simplifies to the Rabi model of quantum optics:

$$H_R = \hbar\omega_c\,\hat{a}^+\,\hat{a} + \frac{1}{2}\hbar\omega_{exc}\,\hat{\sigma}_z + H_{int}$$
$$H_{int} = \hbar g_1\,(\hat{a}\,\hat{\sigma}_+ + \hat{a}^+\hat{\sigma}_-) + \hbar g_2\,(\hat{a}\,\hat{\sigma}_- + \hat{a}^+\hat{\sigma}_+) \quad (2)$$

Here $\hat{a}^+$ and $\hat{a}$ are the creation and annihilation operators of the cavity mode and $\hat{\sigma}_-=|0\rangle\langle 1|$ and $\hat{\sigma}_+=|1\rangle\langle 0|$ are the lowering and raising operators between the ground ($|0\rangle$) and excited ($|1\rangle$) states of the approximated two-level system. The parameters $g_1$ and $g_2$ denote general light-matter coupling strengths; provided the model is derived from Eq. (1), then $g_1 = g_2 = g$. It is important to note that, in contrast to the terms weighted by $g_1$ in Eq. (2), the terms weighted by $g_2$ do not conserve the total number of excitations in the system (as processes of simultaneous photon creation and atom excitation are possible). The latter terms are referred to as anti-resonant or counter-rotating and can be omitted when the light and matter frequencies are close and the coupling between the atom and the cavity is much smaller than the resonance frequencies: this is the so-called rotating-wave approximation (RWA). For not-too-strong light-matter coupling (i.e., $g_1 \ll \omega_c, \omega_{exc}$), the RWA provides satisfying results, but for stronger couplings, large frequency mismatch or multi-photon processes, the RWA breaks down. Then, the counter-rotating terms become relevant. These should be included in any theory that aims to deliver a non-perturbative real-space treatment of electrons, ions and photons.

*Jaynes-Cummings model.* The RWA reduces the Hamiltonian of Eq. (2) to the Jaynes-Cummings (JC) model ($g_2 = 0$), where the excitation number is a conserved quantity. The JC spectrum is a ladder of dressed polaritonic states (see diagram in this Box) with an energy difference that scales nonlinearly as $\sqrt{n}$, where n is the total number of quanta in the system (JC ladder). The extensions of the Rabi and JC models to multiple two-level systems correspond to the Dicke and Tavis–Cummings models, respectively[52].

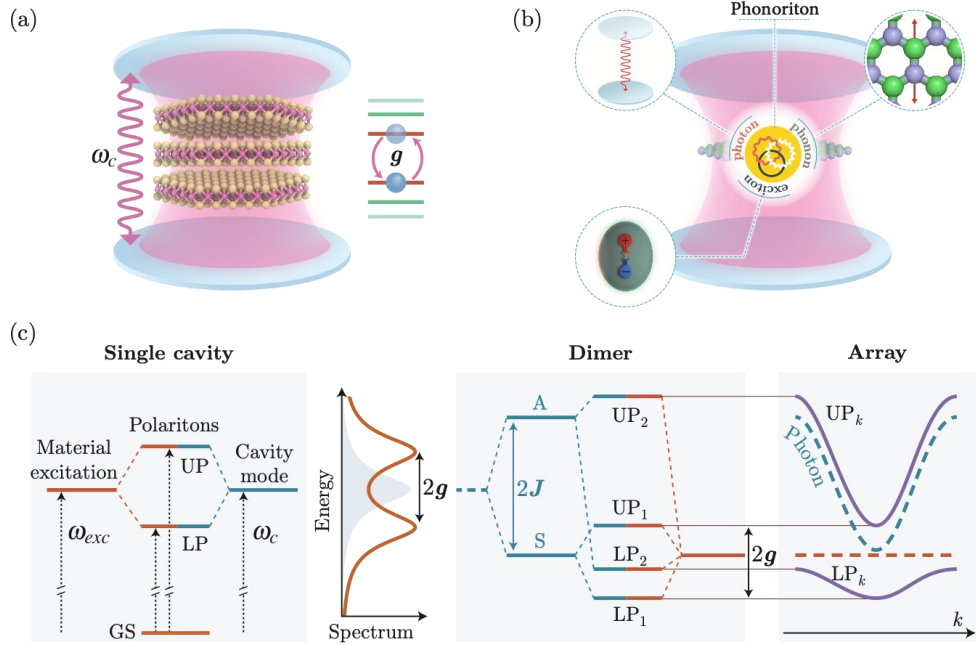

**Models of light-matter interaction and dressing.** Panel A: Quantum material integrated in a cavity. Panel B: The realization of a higher-order cavity-polariton state (a phonoriton[139]) involving multiple degrees of freedom (hybridization between excitons, phonons, and photons). Panel C: The JC model originates from the Rabi model assuming the RWA approximation, and yields a spectrum that can be analytically solved. The vacuum Rabi splitting, proportional to twice the light-matter coupling g, $\Omega_R=2g$, is also indicated. Distinct dressed states' upper and lower polaritons (UP, LP) in a single cavity or in a cavity dimer evolve into polaritonic bands in an array (right panel).

2E. Polaritons in circuit QED

Artificial atoms in circuit QED also give rise to polaritons. In particular, a superconducting qubit – an LC circuit with a Josephson junction producing anharmonicity in the energy spectrum – behaves as a two-level system coupled to microwave cavity modes. If only one resonant mode is important, this system can be modeled with a JC Hamiltonian (Box 2), displaying the expected vacuum Rabi splitting[278–280]. Note, however, that in many cases, the large coupling between the qubit and the cavity prevents a simple description in terms of a single optical mode. Instead, one needs to include coupling to far off-resonant modes to correctly describe the Lamb shift and decay rate of the qubit due to the coupling with the environment. Circuit QED polaritons have been observed in different types of photonic structures (Fig.1C), including (microwave) photonic crystals. Provided the qubit resonance frequency lies in the photonic bandgap, a polaritonic bound state is formed where light becomes spatially trapped near the qubit[281,282]. The spatial confinement of this bound state scales with the detuning between the qubit resonance frequency and the photonic crystal cavity. When multiple qubits are coupled to the same photonic crystal, controlled detuning can be used to engineer coherent interactions among the qubits, enabling quantum simulation of diverse spin models[11].

Superconducting qubits coupled to open transmission lines can emulate cavity QED setups, where the mirrors in the cavity are not photonic structures but (arrays of) qubits themselves. When a central ancilla qubit is excited, coherent population transfer between the ancilla and the "mirror" qubits ensues. The cooperativity of this system can be made very large (above $10^2$) by combining the large Purcell enhancement experienced by qubits close to the transmission line ($\sim 10^2$ times larger than their damping and dephasing rates) with the control over relative positions of the qubits. Circuit QED systems are well suited to achieve an ultra-strong coupling regime[283].

**SECTION 3: Slow light, topology and synthetic quantum matter**

Polaritons offer versatile solutions for creating correlated and strongly interacting light-matter states. The tunability of both the light and the matter constituents propels the use of interacting polaritons as building blocks of complex topological systems and synthetic quantum matter.

3A. Electromagnetically induced transparency (EIT) and slow polaritons

A spectacular example of polariton physics is that of dynamics under EIT conditions, where a "transparency window" can be opened in an otherwise optically dense medium (Fig.7A,B). Within the EIT medium, photons travel as coupled excitations of light and matter called dark-state polaritons (Section 2D). Like many other polaritons, EIT polaritons inherit interactions from their matter component. The physics of EIT is captured by a three-level Lambda ($\Lambda$) system (insets in Fig.7A,B). The properties of the optical field that couples the ground state $|1>$ to an optically excited state $|3>$ can be altered by applying a "control" field tuned near resonance with the transition $|3>$ to $|2>$. The combined effect of the two applied fields is to create an optically dark state, a coherent superposition of states $|1>$ and $|2>$, via destructive interference between the two excitation pathways. Because the transitions to the excited state $|3>$ are canceled, one attains vanishing absorption. The creation of optically dark states in a solid is commonly referred to as coherent population trapping[284] (Fig.7A).

The properties of EIT polaritons can be regulated by the control field. Notably, the group velocity of polaritons traveling in the "transparent" medium under EIT conditions can be arbitrarily reduced[66,285]. Ultra-slow polaritons become increasingly matter-like and quantum information originally carried by photons is stored in long-lived states of the polaritonic medium. The procedures of storing and retrieving excitations via the EIT process constitute the backbone of quantum memories, which are a crucial element for quantum networks.

The EIT physics has been extensively studied in atomic clouds and has also been realized with superconducting qubits[286] as exemplified in Fig.7B. We remark that non-interacting, optically controlled EIT can be realized in purely classical systems. Two oscillators with judiciously chosen frequencies, oscillator strength and damping rates allow one to attain the large variation in the (effective) refractive index that prompts a narrow transmission window. This physics has been explored in various solid-state settings, including optical[287] and plasmonic resonators[288] and graphene plasmons[289]. An intriguing theoretical proposal to achieve EIT involves magnon polaritons in yttrium iron garnet (YIG) integrated in an optical cavity[290].

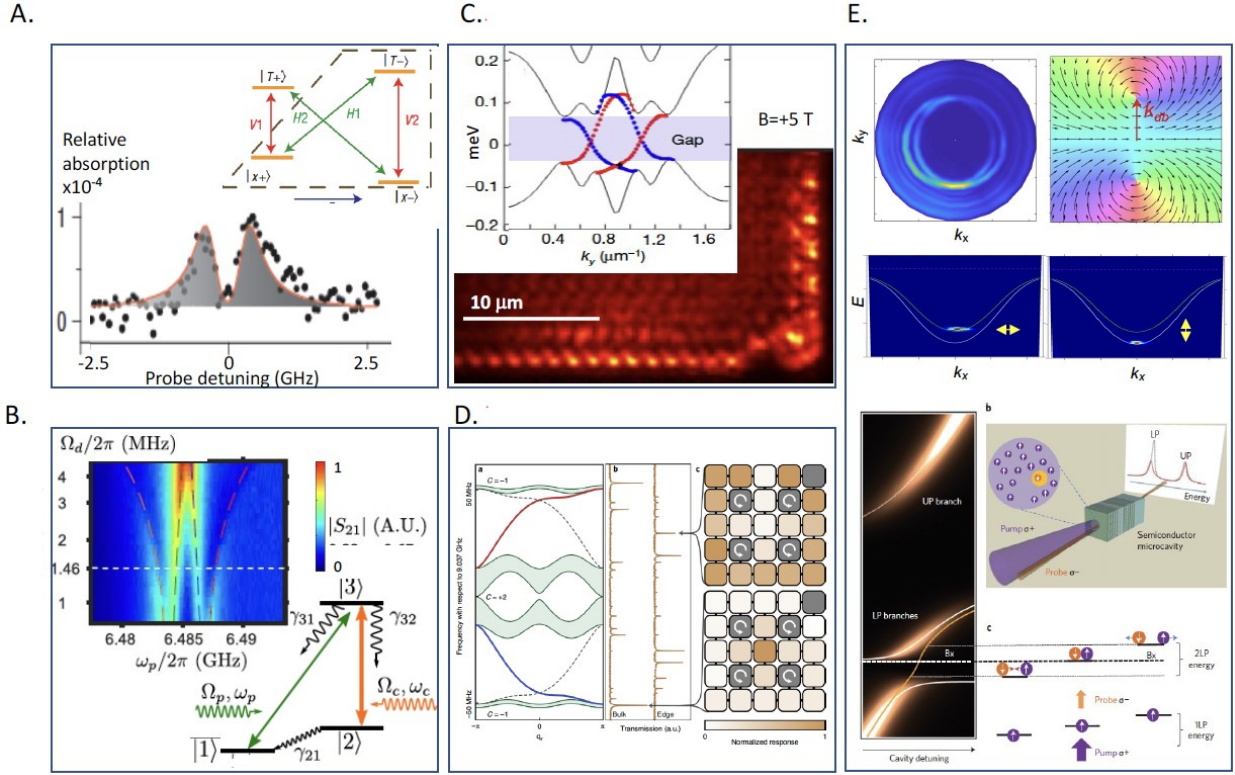

**Figure 7. Electromagnetically induced transparency, topology and phase transitions in polaritonic systems.** Panel A: Coherent population trapping in InAs quantum dots enabling electromagnetically induced transparency manifest by reduced absorption at zero probe detuning. Adapted from Ref.[284]. Panel B: Electromagnetically induced transparency in circuit QED. Four polariton states prompted by the coupling of a superconducting qubit to a resonator are visible in the transmission spectra. Three of the four polariton states form a Λ-system composed of an excited state connected with two ground or long-lived metastable states, where the probe responds to a control (c) field. Adapted from Ref.[286]. Panels C and D: Light-matter interfaces giving rise to topological electrodynamics. Panel C: Exciton-polaritons in semiconductor microcavities. Photoluminescence (PL) from an exciton-polariton condensate reveals a protected topological edge mode. The inset displays the topological band structure of the polaritonic honeycomb lattice with a gap prompted by Zeeman splitting. The structure in the main panel supports the unidirectional flow of a polariton wavepacket around the edge of the array. Unidirectional modes within the topological gap are shown in the inset in red (right-moving) and blue (left-moving). Adapted from Ref.[291]. Panel D: A superconducting Chern circuit. The left panel displays the numerically computed band structure of this model (as implemented) for an infinite strip geometry. Microwave transmission spectra are measured through the actual 5x5 lattice (right), where both the bulk bands and the chiral edges manifest as well-resolved modes due to the finite system size. Adapted from Ref.[292]. Panel E (top): An example of synthetic quantum matter in a condensate of spin-orbit coupled (SOC) exciton-polaritons formed in an optical cavity hosting anisotropic $CsPbBr_3$ microcrystals. The top left panel shows 2D momentum mapping at 574.5 nm. The two SOC polaritons produce offset rings, with spin (polarization) textures shown on the right. Dispersion plots ($E,k_x$) display SOC-split exciton-polariton states driven to the condensates with orthogonal polarizations. Adapted from Ref.[293]. Panel E (bottom): Feshbach resonance with exciton-polaritons in a semiconductor microcavity. The energy of the spin-down exciton-polariton is plotted as a function of the cavity detuning in the presence of a spin-up polariton population. In the perturbative regime, the lower polariton resonance shows a dispersive shape (orange line) around the crossing point with the biexciton energy state (bold dashed line). At higher polariton spin-up densities, the lower polariton branch splits into two at the biexciton crossing energy (solid white lines). This splitting results from the strong

coupling with the biexciton resonance. Upper polariton and lower polariton (UP and LP) branches stand for the upper and lower polariton resonances, respectively. Right: a schematic of the microcavity experiment. Adapted from Ref.[294].

### 3B. Apropos of topology and polaritons

Topological photonics and polaritonics draw inspiration from concepts conventionally explored in solid-state physics, finding analogs in the optical domain. In the Landau paradigm, phase transitions are described by local order parameters. Topological order departs from this paradigm, and one needs to invoke the mathematics of topological invariants.

*Symmetry considerations for topological polaritons.* In their landmark work, Haldane and Raghu[295] predicted the counterpart of this paradigmatic state of matter in photonic crystals with broken time-reversal (TR) symmetry. Photonic systems submit to three general approaches to TR breaking: i) applied magnetic fields or gyromagnetic elements incorporated in photonic structures, ii) synthetic magnetic fields introducing the effective magnetic flux for charge-neutral photons or polaritons (Section 3C), and iii) time-varying optical excitation (Floquet engineering) employing proper driving protocols[296]. All three methods have also been utilized for creating topological circuits in the microwave domain. Methods ii) and iii) are being harnessed for the experimental implementation of topological Bloch bands formed in optical lattices of neutral ultracold atoms[297,298].

*Edge states in topological systems.* A hallmark of topological phenomena is the emergence of edge states at the boundary between regions with distinct topological invariants[296,299–301]. Gapless edge states form within the energy gap of the surrounding media and enable robust unidirectional transport. Edge states comply with the bulk-boundary correspondence: a fundamental principle applicable to both bosonic and fermionic systems. According to this principle, the number of chiral edge modes is uniquely determined by the topological invariant of the bulk. Here, we highlight polaritonic edge modes detected in a honeycomb lattice of coupled semiconductor microcavities (Fig.7C). The dispersion relation due to the honeycomb structure named "polaritonic graphene" displays Dirac cones. Unlike physical graphene, where the spin–orbit interaction is extremely small, "polaritonic graphene" reveals a sizable gap in the presence of an applied magnetic field. In a zero magnetic field, the two central parabolas touch each other at the Dirac cones of the honeycomb lattice. The degeneracy at the crossing points is lifted by Zeeman splitting. The net effect is that Dirac cones transform into four inverted parabolas with chiral edge states residing in the gap regions (insets of Fig. 7C). The dispersion relations in Fig. 7C featuring gapped bands and intragap edge states are analogous to the electronic band structure of a topological insulator[44,302,303]. The photoluminescence exploration of the exciton-polariton dynamics in Fig.7C was performed by populating the chiral edge mode via a polariton condensation mechanism[291]. The applied magnetic field also provides a TR symmetry-breaking mechanism. Data in Fig.7D revealed the possibility of a polaritonic quantum Hall state. Edge states have been documented in polaritonic media implemented in exciton-polariton[304,305] and circuit QED[292,306] platforms.

*How to imbue polaritonic quasiparticles with topology.* Polaritons are theoretically anticipated to acquire topological properties from their matter constituents. Specifically, topological electronic states in Weyl semimetals[307] and topological insulators[308] are predicted to produce topologically nontrivial plasmon polaritons. Experiments probing such candidate systems have been attempted[75,309–311] but have yet to reveal topologically nontrivial polaritonic properties. Alternatively, polaritonic quasiparticles can also inherit topological properties from their photonic constituent. For example, topological photonic states can be engineered in chiral cavities[312] and in dielectric waveguides (photonic crystals) without resorting to external magnetic fields or to gyromagnetic elements[313]. Then the strong coupling between cavity topological photons and material excitations gives rise to topological polaritons. This latter approach was utilized to demonstrate both topological exciton-polaritons[44,314] and topological phonon polaritons[181], both revealing edge modes. Experiments reported in Refs.[44,181,314] are polaritonic implementations of the quantum spin Hall effect and rely on topological properties born out of lattice symmetries, similar to crystalline topological insulators in solids.

*Topological phase transitions.* Graphene/$MoO_3$ heterostructures host higher order plasmon-phonon polaritons with the hyperbolic phonon constituent. These hybrid polaritons undergo a topological transition, where the isofrequency contour transforms from an open to a closed shape depending on the doping level of graphene[250,315]. Right at the topological transition, the contour acquires a quasi-flat shape for a broad range of wavenumber distributions. It was theoretically proposed that free-electron−photon interactions in the heterostructure can be enhanced due to this quasi-flat isofrequency surface region[316].

*Topological textures.* We conclude this section by mentioning a class of topological polaritonic effects pertaining to real-space defects, vortices and textures[86,317]. For example, advanced nano-imaging experiments have been utilized to read out topological charges in plasmon polariton[46,86,318], phonon polariton[46,230,251,315] and hybrid plasmon-phonon polariton[250,315] platforms.

## 3C. Bose-Einstein condensates of polaritons

Bose-Einstein condensates (BECs) are spectacular examples of macroscopic quantum behavior in matter. Boson condensation was experimentally realized in ultracold atom gases in 1995. Following this success in cold atoms, condensation of exciton-polaritons was predicted[319] and demonstrated in microcavities containing semiconductor quantum-well systems[107,320–322]. Since exciton-polaritons are much lighter than atoms (exciton-polariton masses can be as small as $10^{-5}$ of the free electron mass $m_e$), their condensation can occur at room temperature in polymers[323], organic semiconductors[324,325] and $CsPbBr_3$ perovskites[326]. Ambient temperatures make exciton-polariton condensates in solids a particularly attractive platform for applications in quantum devices[327,328]. Apart from exciton-polaritons, plasmon polaritons in metallic micro-structures also reveal BECs[329].

Due to the relatively short lifetimes of exciton-polaritons, their condensates do not reach complete thermal equilibrium and therefore are not standard BEC states. A notable exception of BEC condensate of polaritons at thermal equilibrium was discussed in Ref.[330]. However, they satisfy the two defining criteria of condensation: i) they exhibit long-range spatial coherence, and ii) this coherence is spontaneous rather than externally driven, as in a laser. The non-equilibrium nature of the exciton-polariton BEC leads to many intriguing phenomena, e.g., pattern formation and modified collective mode spectra. A comprehensive review on the topic of exciton-polariton condensates is due to Kavokin et al.[107].

Non-equilibrium aspects aside, the ground state of exciton-polaritons is theoretically predicted to exhibit a rich phase diagram, changing from excitonic to photonic condensate as a function of particle density and cavity detuning[331]. Having unequal electron and hole densities in the microcavity is another pathway to realizing new types of quantum matter. Such a density imbalance introduces frustration in the system, which may result in the formation of condensates with spontaneously broken time-reversal or rotational symmetries[332].

## 3D. Polaritonic simulators and synthetic quantum matter

Polaritonic systems excel in emulating electronic phenomena in quantum materials and also host novel effects rooted in the dual light-matter nature of polaritons.

*Synthetic magnetic fields.* Polaritonic effects that stem from synthetic magnetic fields offer illustrative examples of emulations. Whereas the influence of ordinary magnetic fields is restricted to charged particles, synthetic gauge (magnetic) fields rooted in Berry curvature can be engineered for a variety of neutral systems ranging from atoms in optical lattices[333,334] to photonic and phononic resonator arrays[296]. Control of the photon component of a polariton opens pathways for generating effective spin-orbit coupling (SOC)[335,336]. In the realm of exciton-polaritons, TE-TM splitting effectively emulates a winding in-plane magnetic field on the photon pseudospin represented by the Stokes vector and in-plane optical anisotropy serves as a constant magnetic field to break time-reversal symmetry[337]. Small in-plane anisotropy for exciton-polaritons in semiconductor quantum wells may be introduced by the optical anisotropy of the cavity[338]. Strong anisotropy and, thus, a prominent synthetic magnetic field can be obtained from semiconductors with intrinsic and strong birefringence[339]. Thus, birefringence imitates the Zeeman splitting of material resonances in the presence of a physical magnetic field (Fig. 7E), as demonstrated for exciton-polaritons in lead halide perovskites[293]. In another example, a one-dimensional array of interacting semiconductor quantum wells, each hosting BECs, can be programmed by photoexcitation to emulate topological nontrivial polaritonic band structures[340].

*Polaritonic simulators with programmable properties.* A goal of (polaritonic) quantum simulations is to advance the still incomplete understanding of even more complex phenomena such as high-$T_c$ superconductivity and strongly correlated physics[341,342]. Notably, moire superlattices discussed in Section 2B are emerging a robust quantum simulation platform that enables the study of strongly correlated physics and topology in quantum materials[192]. Polaritonic quantum simulators provide a unique toolset: readily programmable dispersion relations and interactions[11,147,343]. Moreover, readout of desired observables can be done optically. Despite their bosonic nature, polaritons reveal a broad range of phenomena that are commonly associated with fermionic transport and the dynamics of electrons in materials. Some of these polaritonic counterparts of electronic properties stem from the universal attributes of waves in periodic potential, e.g., Bloch oscillations (BOs) and polaritonic band structures. Other effects, including quantum blockade (Section 4), originate from interactions that polaritonic quasiparticles acquire from their matter components. To avoid unnecessary redundancy with comprehensive articles on polaritonic simulators[11,106,147,279,343,344], in the remaining paragraph of this Section we focus only on selected examples of simulations.

*Polaritonic Feshbach resonance.* A less commonly described simulator is a polaritonic Feshbach resonance. The central tenet of Feshbach physics is that nearly stable bound states of two quantum mechanical objects (resonances) govern how these two objects interact in the regime where their total energy is in the vicinity of the bound state. Provided the total energy of the objects is approaching the resonances, the interaction of the objects is enhanced and can be switched from repulsive to attractive. The ability to manipulate both the sign and the strength of interactions underlies many phenomena in AMO physics, including fermionic superfluids and the generation of ultracold molecules, and is central to the realization of quantum simulators. Exciton-polaritons in microcavities also display Feshbach interaction (Fig.7E, bottom)[294]. Exciton-polaritons acquire a polarization degree of freedom that prompts distinct matter content in the states associated with the two circular polarizations of their photonic part. By tuning the energy of two polaritons borne out of two counterclockwise circular polarizations across the biexciton bound-state energy, it becomes possible to enhance attractive interactions.

*Synthetic exciton-polaritons.* Our second example of a less common simulator is an experimental implementation of a synthetic exciton-polariton realized on an ultracold atom platform[345]. In these experiments with an optical lattice, the exciton is effectively "replaced" by an atomic excitation, whereas an atomic matter wave emulates the photon. The two constituents hybridize and produce a dispersion relation mimicking that of the exciton-polaritons. Our next example is semi-Dirac quasiparticles. Conventional 2D fermions are described by parabolic energy ($E$)-momentum ($k$) dispersion $E(\mathbf{k}) = \hbar^2 k^2/(2m)$ with an effective mass $m$. In contrast, the Dirac fermions have linear dispersion $E(k) = \hbar v_F k$ and are massless. Semi-Dirac fermions appear in materials where two Dirac points merge into a semi-Dirac point[346]. Prior to the discovery of semi-Dirac fermions in the electronic system of ZrSiSe, the semi-Dirac dispersions had been experimentally explored only in synthetic platforms, including honeycomb lattices of ultracold atoms[347], photonic resonators[348,349], and exciton polaritons in photonic crystals[350].

*Polaritonic Bloch oscillations.* We recall that electrons accelerated in a crystal potential undergo a periodic motion known as Bloch oscillation (BO). Polaritonic platforms have been utilized to emulate BOs. These oscillations are a product of coherent wave dynamics in a lattice subjected to the DC or slowly varying field; therefore, BO may be realized for other forms of propagating waves. In experiments with plasmon- and exciton-polaritons, a gradient of the effective refractive index emulates the linear potential associated with the electric field in the electronic version of BOs ([351,352]). Notably, polaritonic BOs can be directly visualized by imaging.

*Polaritonic drag.* In condensed matter systems, quasiparticle flow commonly instigates a transfer of energy and momentum between separate but interacting subsystems of quasiparticles. These effects are commonly categorized under the notion of drag. Drag of plasmon polaritons by a DC current was observed in graphene[31,32]. Plasmonic nano-imaging data revealed that the concept of drag extends to the two constituents of the same polaritonic quasiparticle: a superposition of infrared photons and Dirac electrons. DC currents solely perturb the electronic constituent of the quasiparticles, and the photonic component reacts according to the rules of quasi-relativistic theory of polaritons in graphene. The magnitude of DC drag is theoretically expected to be enhanced, provided the polaritons are slowed down in media with properly engineered dispersions.

## 3E. Nonlinear polaritonics in quantum materials

Nonlinear polaritonics is an emerging area with great potential impact. As polariton waves are electromagnetic in nature, nonlinear polaritonics has many common traits with the familiar field of nonlinear optics, although much enriched by quantum phenomena that underpin the matter components of the polaritonic platforms in Figs.1,3. While the wave character of a polariton ensures longer-range interactions than for most other quasiparticles (Section 1), the matter component often further boosts these interactions. Nonlinear coefficients may therefore be far larger than what is observed in conventional nonlinear optics. Nevertheless, examples of nonlinear polaritonics are still sparse[353]. A broad class of nonlinear polaritonic effects pertain to quantum nonlinearities in the single-photon regime, which we discuss in Section 4. Here, we focus on nonlinear effects prompted by intense optical fields, in which the electromagnetic component of the polariton is classical and all quantum effects are associated with the matter constituent. Polaritonic enhancement of weak optical nonlinearities has been systematically explored in plasmonic nanostructures[354–357]. Furthermore, polaritonic coupling relaxes symmetry constraints that govern higher-order susceptibilities and thus enables prominent higher-order harmonic generation[355,358–361].

*Strong optical fields and driven systems.* An active area of research covers nonlinear phonon polaritons, which connect THz nonlinear optics with structural phase transitions. In the linear regime, phonon polaritons are well understood as associated with the response of polar crystals to THz and infrared radiation[362]. Nonlinear attributes of phonon polaritons have historically been explored mostly in experimental platforms involving impulsive stimulated Raman scattering excitation with propagating near-infrared pulses[124,363–365]. In these studies, steering, focusing, and amplification of THz radiation occur in regions of the crystal beyond those traversed by the excitation pulse[366,367]. However, even in cases in which phase matching was optimized to drive phonon polaritons of the largest amplitudes, experiments based on near-infrared laser pulses and stimulated Raman coupling never reached the deep nonlinear regime necessary to manipulate the structural properties of materials and to initiate phase transitions. Subsequent work combined the accumulated knowledge from these studies with advances in direct nonlinear phonon excitations[368]. Phonon polaritons were excited to far larger amplitudes than those achievable by impulsive Raman means. In these latter experiments, the nonlinear response was, for example, shown to yield light-induced switching of the ferroelectric polarization[369,370]. The propagating nature of phonon polaritons gave access to nonlocal nonlinear electrodynamics by spatially separating the functional response and the optical drive over macroscopic distances. Furthermore, higher-order harmonics at the multiples of phonon-polariton frequencies were also detected in this regime, enabling mapping of the microscopic interatomic potentials[371].

*Dynamics of driven ferroelectrics.* As the propagating nature of phonon polaritons appears to permeate the physics of polarization switching in ferroelectric $LiNbO_3$, it is not unlikely that polaritonic effects may also play a role in recent experiments in which mid-infrared[372] and THz[373] pulses were used to induce long-range ferroelectric order in the quantum paraelectric $SrTiO_3$. Although there is no explicit discussion of polaritonic effects in the existing literature on photo-induced ferroelectricity, this is an interesting avenue for future research, especially because these phenomena are intimately connected to the possible effects of polariton-induced renormalization of fluctuations near a quantum phase transition[374].

*Nonlinear phonon polaritons* have been studied in conjunction with other excitations and have relied on spin-lattice interactions to generate mixed modes of the phonon-magnon polariton type, typically in magnetoelectrics and in multiferroic materials[375,376]. For example, optical excitation of phonon polaritons with distinct frequencies in an anisotropic crystal of $ErFeO_3$ allows one to generate strong pseudo-magnetic fields[130]. The interlayer dipolar exciton-polaritons in the bilayer vdW material $MoS_2$ display unprecedented nonlinear interaction strengths prompted by strong interactions between polaritonic quasiparticles[377].

*Nonlinearities in superconductors.* We conclude this section by briefly mentioning Josephson plasma waves in layered superconductors, which have been shown to prompt strong polaritonic nonlinearities in the THz frequency range[378]. These modes are rooted in Cooper pair tunnelling between $CuO_2$ layers in cuprate superconductors and are endowed with extremely large, tunable nonlinearities that are related to the response near critical current and phase slip regimes. Propagating Josephson plasma polaritons supported by interlayer superconducting pair tunnelling have been predicted[379] and shown experimentally to give rise to exceptionally large nonlinearities[380] as well as to interference effects prompting electromagnetically induced transparency[381]. Additional theoretical work has recently been reported in this area, exploring the generalized electromagnetic response in both linear[382] and nonlinear[383] response regimes.

## 3F. Magneto-polaritons in quantum materials

Magneto-polaritons form when light interacts with magnetic dipoles (e.g., magnon polaritons) or electric dipoles that are coupled to the magnetic field (e.g., magnetoplasmon or magnetoexciton polaritons). Such polariton waves in magnetic fields exhibit many intriguing properties, including tunable light confinement[384], polarization conversion[385,386], and nonreciprocal propagation $(\omega(\vec{k}) \neq \omega(-\vec{k}))$[387–392]. For instance, chiral and unidirectional polariton waves that travel at sample edges are predicted to be backscattering-immune[303,393] and can be topologically analogous to the 2D topological superconductor with chiral Majorana edge states and zero modes[300]. Their low damping may serve as the key ingredient for low-energy-consumption on-chip photonic devices that utilize strong light-matter interactions. Indeed, studies of the nonreciprocity of surface magnon polaritons in magnetic materials[394–396] and nonreciprocal plasmas (e.g., edge magnetoplasmons) in semiconductors or 2D electron gases[397–400] date back to the 1960s-1980s. However, most of these pioneering works were limited to theoretical studies or were performed at microwave frequencies due to the lack of appropriate real-space examination techniques.

*On magneto-nano-optical experiments with polaritons.* One notable recent advance in detecting magneto-polaritons at infrared frequencies is the use of magneto-SNOM[143,144,401] to study Dirac magnetoexcitons in near charge-neutral graphene[143,144,401,402]. Here, the magnetoplasmons and magnetoexcitons are directly associated with Landau level (LL) transitions in a magnetic field – specifically, with the transitions between adjacent LLs. The polariton wavelength $\lambda(\omega) \simeq \frac{g(\omega)}{2\pi} \frac{v_F}{(\omega-\omega_c)}$ is determined by the cyclotron resonance $\omega_c$[144,403]. Here, $g(\omega)$ is a dimensionless coupling parameter $g(\omega) = e^2/\hbar \kappa v_F \approx 1$ and $\kappa(\omega) = \kappa_1 + i\kappa_2$ is the effective dielectric permittivity of the graphene environment. The group velocity of the Dirac magnetoexcitons is usually small and can be close to the Fermi velocity $v_F$. This small group velocity, which differs from the velocity in previously studied plasmons in electric-gated graphene, might lead to stronger light-matter interactions. This type of magneto-polariton is inherently induced by Landau level transitions and is therefore likely to inherit important edge properties. Future magneto-SNOM experiments need to address this hypothesis of edge magneto-polaritons.

*Novel controls of polaritonic effects.* Magneto-polaritons may unlock new pathways for the control and manipulation of confined light-matter interactions using magnetic fields. For instance, phonon polaritons in hBN (Fig. 4 B) can be completely shut down via the coupling to magnetoexciton polaritons in an adjacent graphene layer in a quantized manner at discrete magnetic fields[143]. The coupling strength can exceed the strong coupling threshold and is, in principle, can be tuned via the layer thickness[37,226]. Other interesting hybrid polariton systems that involve magnetic interactions include magnon-phonon polaritons in a ferroelectric (LiNbO$_3$)-antiferromagnetic (ErFeO$_3$) heterostructure[375], plasmon-magnon excitations in a graphene-2D ferromagnet CrI$_3$[404], and magnon-exciton polariton coupling in microcavities hosting anisotropic layered semiconductors including CrSBr[163,405]. The ability to engineer hybrid polaritonic states across diverse materials such as 2D magnets uncovered the potential for understanding light-matter interactions in new regimes.

## 3G. Polaritonic losses and loss mitigation

Polaritonic virtues are appealing, yet their inherent losses are appalling. This is a common refrain of numerous works exploring the many aspects of polaritons. Both ohmic or material losses and photonic losses contribute to the dissipation of polaritons. In this Section we discuss an approach to reducing photonic losses by harnessing the special properties of bound states in the continuum[406] (BIC) and another approach to mitigating material losses using the concept of "virtual gain"[407].

*BIC polaritonics.* Photonic BIC offer a novel approach for creating polaritonic states. BIC modes are symmetry-protected against coupling to the radiative field, prompting resonances with the vanishing spectral linewidth and infinite quality factors[406]. Photonic BIC can be readily achieved in periodic structures and photonic crystals[408]. Dielectric gratings supporting BIC modes promote strong light-matter coupling quantified by the Rabi splitting of 50-70 meV even in the case of monolayer vdW semiconductors[409,410]. Exciton-polaritons in BIC-assisted structures reveal narrow polaritonic widths. BIC-enabled long polaritonic lifetimes promote Bose-Einstein condensation at extremely low excitation thresholds and also promote large nonlinear interactions[328]. Weber et al. fabricated self-hybridized BIC structures based on $WS_2$ semiconductors; these latter structures reveal a Rabi splitting as high as 116 meV along with a quality factor that is limited by material resonances[411]. The study of plasmonic BIC is likewise an interesting research direction. For example, one-dimensional BIC gratings made of silver produced quality factors exceeding by an order of magnitude those of ordinary metallic structures[412].

*Mitigating losses with synthetic frequencies.* Ohmic losses are ingrained in the scenario of a photon dressed in a material excitation as in Fig.8A. Ohmic losses reduce the magnitude of wavevectors attainable with polaritons, limit the resolution of polaritonic imaging and restrict the propagation range of polaritonic wavepackets (Fig.8C, top), prompting the search for loss-mitigating solutions. One way to counteract the losses is to introduce an optical gain. Gain-embedded magnonic platforms displayed polariton-based microwave amplification with quality factors exceeding $Q>10^9$[413]. In the case of plasmonics, a gain-based solution requires challenging materials engineering[414–416]. An artificial or virtual gain[407] enabled by temporal shaping of optical pulses and synthetic frequencies presents an enticing alternative. Indeed, the complex frequency wave $\omega=\omega_0 - i\Gamma/2$, where $\Gamma$ is the scattering rate, leads to the elimination of the imaginary part of the dielectric function, which stands for the ohmic dissipation. The precise choice of complex wave in the case of polaritonic propagation $\omega = \omega_0 - i\beta$ is more nuanced since the requisite decay parameter $\beta$ depends on the central frequency $\omega_0$ but its linewidth is generally of the order of the resonance mode in Eq.1. Other nuances of polaritonic effects activated by synthetic frequencies were discussed in Refs.[417–420]. The authors of Ref. [421], following protocols proposed in Ref.[422], emulated the polariton excitation with a complex field $\omega_0 - i\beta$ constructed from a linear combination of multiple real frequencies (Fig.1B). When applied to phonon polaritons in $MoO_3$, this analysis revealed the extend propagation range. It was proposed in Ref.[423] that broad-band pulses produced via difference frequency generation (DFG) methods and routinely used in polaritonic imaging may possess the time structure required to obtain the virtual gain. This latter proposal awaits direct experimental tests.

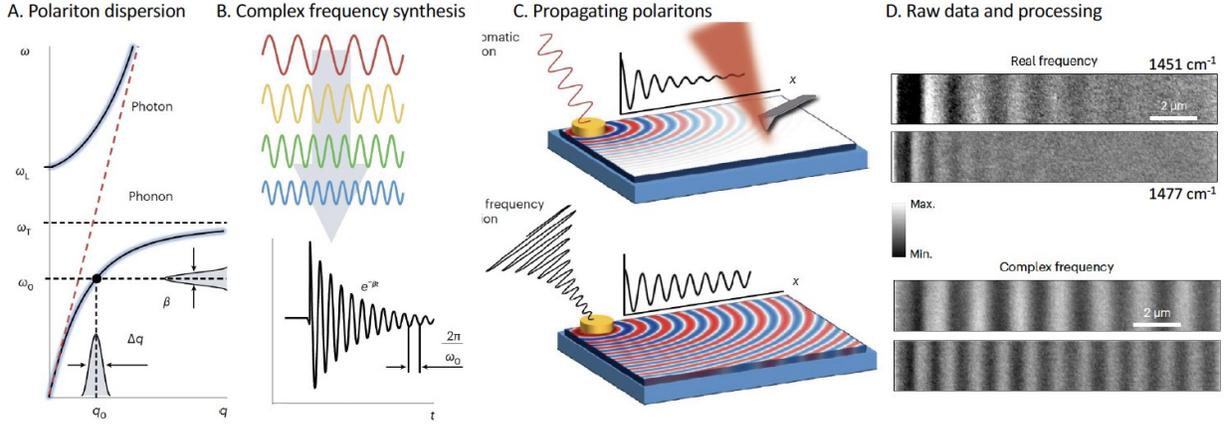

**Figure 8.** Synthetic frequencies and their benefits for polaritonic imaging. Panel A: The coupling of a photon and a dipole-active resonance is exemplified for an infrared active phonon and leads to the characteristic mode repulsion and energy partitioning into two branches of the frequency-momentum ($\omega,q$) dispersion. A polaritonic gap is formed between the transverse and longitudinal optical frequencies $\omega_T$ and $\omega_L$. Panel B: Complex frequency waves can be synthesized from monochromatic waves. The complex frequency wave is necessarily attenuated in time, a condition enforced by causality. Panel C: Polaritonic waves emanating from a circular launcher. Insets schematically display the line cuts. Even in the absence of ohmic dissipation, polaritonic fringes emanating from a circular launcher decay because of the geometric loss. Panels A-C adapted from Ref.[423]. Panel D: Propagating phonon polaritons in $MoO_3$. Experimental data are plotted in the top panels and complex frequency analysis results are presented in the lower panels. Adapted from Ref.[421].

## SECTION 4: Quantum polaritonics

The quantum nature of polaritons reveals itself in multiple ways. First, coherent and linear light-matter interactions preserve the quantum state of photons after hybridizing with matter. Second, in the nonlinear regime, the interaction with matter is mapped to an effective photon-photon interaction. Quantum polaritonics may eventually provide unique resources for quantum information applications.

### 4A. Quantum light creates quantum polaritons

In the linear regime, photons can be coherently mapped in and out from a material excitation while preserving the initial quantum state. For example, single photons can be converted into dark polaritons in a dense atomic medium via electromagnetically induced transparency EIT (Section 3A). Coherent population transfer between quantum states has been reported for microcavity exciton-polaritons[424] and plasmon polaritons produced with either polarization- or energy-entangled photons[425,426]. In all these experiments utilizing diverse solid-state platforms, polaritons inherit the quantum statistics of the photons that generate them. Furthermore, propagating single exciton-polaritons display self-interference[427], documented by direct imaging in Fig.9A. In this experiment, single photons emitted by a quantum dot were utilized to generate, inject and propagate individual microcavity exciton-polaritons. Single-plasmon-polariton interference was observed for polaritonic quasiparticles propagating along a planar metal-air interface[428].

Two-photon interference, as in the canonical Hong-Ou-Mandel (HOM) effect[52] illustrated in Fig.2B, can be used as a reliable probe of the quantum statistics of the particles. HOM experiments have been utilized to inquire into the quantum properties of polaritons[429], with representative results displayed in Fig.9B. HOM-based inquiries into the quantum properties of plasmon polaritons in metallic nanostructures have emerged as an informative method. Plasmonic HOM experiments confirmed that quantum plasmon polaritons (those produced by twin photon pairs) remain indistinguishable even after propagation in lossy metallic structures[430–432]. Quantum state tomography was implemented in the setting of plasmon-polaritonic HOM platforms and was used to quantify the degree of entanglement via concurrence measurements[430]. Notably, losses in polaritonic beamsplitters play an important role in quantum plasmonics. In the presence of losses, a plasmonic HOM platform may yield not only bunching output characteristic of bosonic particles but also unanticipated anti-bunched two-plasmon quantum interference. It has been proposed that quantum interference pathways in the plasmonic setting can be regulated by the judicious design of the reflectance/transmission properties of polaritonic beamsplitters[433]. Progress with single-photon detectors fabricated from atomically layered quantum materials[434] may eventually enable monolithic circuits with on-chip emitters, detectors and polaritonic waveguides, all operating in quantum regimes.

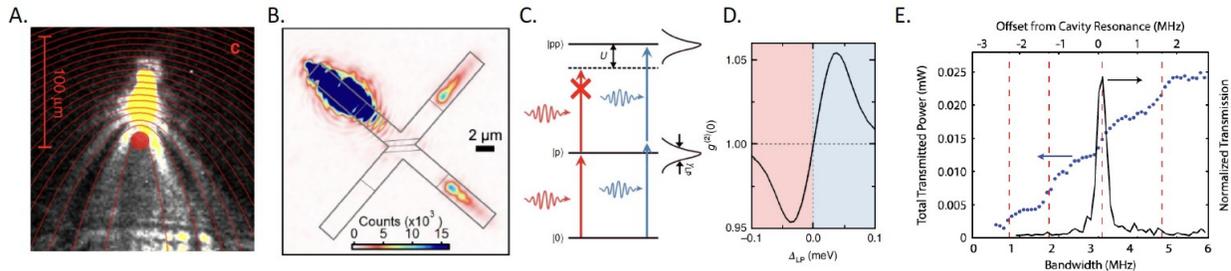

**Figure 9. Quantum polaritonic effects in excitonic, plasmonic and circuit QED platforms.** Panels A, B reveal how single- and two-polariton interference may be used to explore polariton coherence. Panel A: Single exciton-polaritons in microcavities can propagate over macroscopic distances (~$10^2$ μm). When traveling across a defect in the microcavity (red dot), a single exciton-polariton shows self-interference[427]. Panel B: Hong-Ou-Mandel (HOM) interference with plasmon polaritons in metals[435]. The indistinguishability of two photons generated by parametric down conversion is preserved after in-coupling and out-coupling from a plasmonic structure. The resulting plasmon polaritons are made to interfere via an X-shaped plasmonic beamsplitter. Panels C, D and E display nonlinearities at the few-photon level. Panels C and D: blockade effects are observed in the excitation spectrum of exciton-polaritons in a microcavity, where polariton polariton interactions shift the two-polariton state energy by an amount $U$. Detuned photons in Panel D (red) are thus prevented from entering the cavity. Adapted from Ref.[436]. Panel E: Nonlinearities due to strong light-matter interaction prompt blockade in a circuit QED setting. In this example of "dispersive blockade," a superconducting qubit couples off-resonantly with a cavity mode, giving rise to a set of levels whose energy depends on how many photons occupy the cavity. This leads to the emergence of a staircase pattern (blue dots), attesting to quantized microwave transmission[437].

### 4B. Polariton blockade

The "blockade" effect describes how the occupation of an energy level by a particle prevents another particle from occupying the same level, in analogy with Pauli's exclusion principle. While Pauli's exclusion relies on the anti-symmetry of fermionic wavefunctions, one can also implement a blockade via repulsive interactions. Polaritonic blockade is analogous to its Coulomb counterpart in electronic systems, which stems from a charging energy relevant to electron transport in nano- and meso-structures[438]. Transport of polaritons follows the same principles once photons are made to be interacting by dressing light with matter. Experimentally, the polariton blockade is documented through the statistics of light, which can display sub-Poissonian and anti-bunched features.

*The physics of polariton blockade is captured by the anharmonicity of the Jaynes-Cummings ladder of eigenstates* (Box 2), which can be observed when interactions overcome dissipation[439]. It is important to note that energies associated with the transition from the ground state to the first Jaynes-Cummings manifold in Box 2 are distinct from transition energies higher up in the ladder. Therefore, the JC system exemplifies the ultimate nonlinear optical device, enabling the observation of photon-photon interactions at the single-photon level[440].

*Polariton blockade platforms.* The JC Hamiltonian is applicable to multiple microscopic scenarios of blockade effects that have been realized in materials[436,441] and circuit QED platforms[442] (Fig. 9). Promising platforms for realizing polariton blockades at the single-photon level sustain matter excitations that are intrinsically strongly interacting, including dipolar excitons[443,444]. The concept of polaritonic blockade has been theoretically extended to 2D quantum materials[445,446] and experimental studies are emerging. One avenue for investigating polaritonic blockade involves photon coincidence measurements made by means of a direct probe of the second-order correlation function $g^2(\tau)$[15,439]. Alternatively, experimental hallmarks of blockade in the dynamics of surface plasmon polaritons[447] and exciton-polaritons in semiconductors can be probed by exploring the polaritonic response under varying excitation intensity[99]. Karnieli et al. analyzed the electron-polariton blockade of free charge particles introduced in cavity QED[448]. Rydberg-exciton polaritons constitute yet another platform for polaritonic blockade discussed in Section 4C.

## 4C. Rydberg exciton-polaritons

An alternative route towards blockade is via highly excited Rydberg states that display large dipole moments, which enable atom-atom interactions at large distances (on the order of microns). Rydberg polaritons inherit interactions from their Rydberg component, resulting in strong photon-photon interactions at low densities. In atomic systems, Rydberg dressing prevents the excitation of two atoms located within a certain spatial region known as the "blockade radius." Rydberg blockade has been thoroughly studied in atomic clouds under the condition of EIT[449], as well as in Fabry-Perot cavities[450], and can be harnessed to realize photon-photon gates[21].

Rydberg exciton-polaritons have been extensively discussed theoretically and have also been experimentally implemented (Fig.10). Wannier-Mott excitons in inorganic semiconductors are described by the hydrogenic model, implying the formation of excitonic Rydberg series. The Rydberg excitonic states form below the band gap $E_g$ of the 3D or 2D host material at energy $E_n^{3D} = E_g - \frac{Ry}{n^2}$ or $E_n^{2D} = E_g - \frac{Ry}{(n-0.5)^2}$, respectively, where Ry is the effective Rydberg constant and $n$ is the principal quantum number. The excitonic Rydberg states with $n \leq 5$ are routinely observed in the transition metal dichalcogenides[451], whereas $Cu_2O$ displays series with $n$ as high as 25 in Ref.[452] and 30 in Ref.[453]. The similarities between Rydberg states in atoms and in semiconductors are many and varied. For example, high-$n$ excitonic Rydberg states reveal enhanced dipole-dipole interaction extending over macroscopic (micron) distances, much like the atomic Rydberg systems. Likewise, Rydberg exciton-polaritons inherit interactions from high-lying Rydberg states. Rydberg exciton-polaritons were observed in $Cu_2O$ crystals integrated in resonator microcavities[104,454] and were also observed in $CsPbBr_3$[95], $ReS_2$ crystals[455], and monolayer $WSe_2$[456]. Enhanced nonlinearities enabled by Rydberg polaritonic states pave the way for strongly correlated light-matter states in a solid-state platform[456,457]. Finally, we remark that the optically driven inter-Rydberg transitions in $WSe_2$ are marked by an extraordinarily high oscillator strength[458], which facilitates the observation of exciton-polaritons in nano-imaging experiments[459]. The ultra-fast nonlinear optical response of $Cu_2O$ crystals revealed the Rydberg blockade of exciton-polaritons[454].

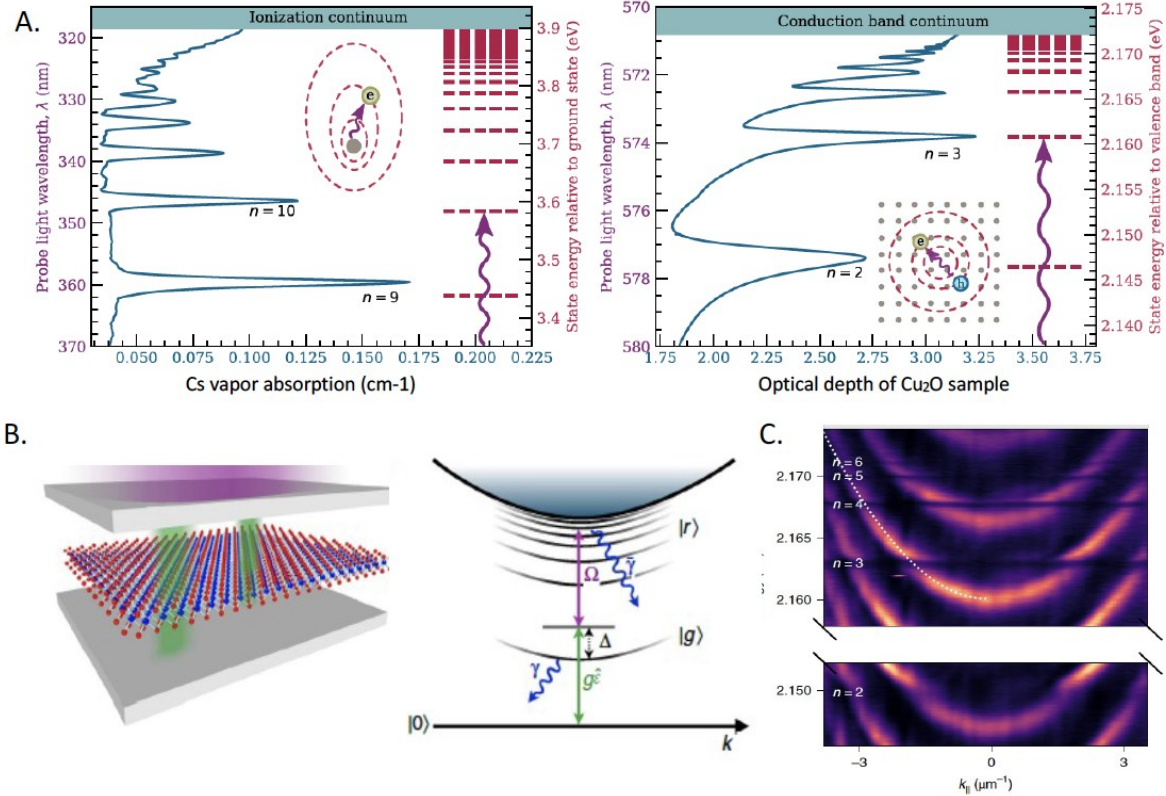

**Figure 10.** Rydberg excitons and exciton-polaritons in atomic and solid-state platforms. Panel A: Rydberg states are ubiquitous in atomic systems and semiconductors. Left: data for Cs vapor; Right: data for Wannier–Mott excitons in $Cu_2O$, reproduced from Ref.[460]. Panel B: In semiconductor microcavities, the cavity field couples to excitons; g is the coupling strength and $\Delta$ is the detuning. The additional field provides for two-photon resonant Rydberg state excitation with Rabi frequency $\Omega_R$. Adapted from Ref.[457]. Panel C: Experimental dispersion of Rydberg exciton-polaritons obtained for the $Cu_2O$ crystal integrated in a resonator cavity with parabolic dispersion[461]. Excitonic resonances ($n$ = 2-6) appear as flat lines, signaling the high effective mass of the excitons. Angle-resolved transmission experiments reveal characteristic avoided crossings along with upper and lower exciton-polariton branches separated by the vacuum Rabi splitting for excitons with $n$ = 3-6.

**Box 3. Light management**

Precisely engineered electromagnetic environments enable the efficient interfacing of photons with material excitations. This Box outlines some of the common design principles for forging polaritonic states with the desired properties and exquisitely controlled interactions. Additional opportunities are enabled by synthetic frequency engineering (Fig.8).

*Cavity encounters.* Optical cavities enhance interactions through improved frequency-dependent matching of the optical field with materials excitation. As they are engineered to tune system-specific interactions, cavity designs are diverse, with dimensions ranging from many free-space wavelengths down to quantum emitter sizes themselves. Photonic cavities constructed from periodically structured materials (cell A) provide a reliable route to enhanced interactions. 2D vdW materials are broadly amenable to cavity integration, giving rise to highly confined surface polaritons with plasmonic, phononic or excitonic matter constituents.

*Dispersion engineering (cells B,C)*. Optical structures such as photonic crystals and metamaterials offer numerous solutions for energy-momentum ($\omega,q$) dispersions engineering[462]. It is feasible to implement optical structures with the energy-momentum ($\omega,q$) dispersions required for promoting negative refraction[463] or flat photonic bands that slow down light and further increase light-matter interactions[464,465]. The 2D Lieb lattices are an example of a two-dimensional array prompting flat electronic and photonic bands. Lieb lattices are being extensively employed in the study of the BEC of exciton-polaritons[466,467] in semiconductor structures. 2D honeycomb photonic structures give rise to linearly dispersing polaritons[468] analogous to the linear dispersion of the electronic bands in honeycomb graphene monolayers. An extreme case of dispersion engineering is the formation of exciton-polaritons with negative effective mass in $WS_2$ crystals embedded in dielectric cavities[469]. Moreover, advanced structures can endow light across the electromagnetic spectrum with chirality and spin- or orbital angular momenta. Customized photons in such cavities enable newfound polaritonic properties and give rise to behaviors emulating topological effects, including unidirectional modes, non-reciprocity, and other attributes discussed in Section 3.

*Plasmonic nano- and pico-cavities* (cells D and E). A reduced mode volume V of cavities, down to the zeptoliter range[35], is desirable for achieving significant light-matter coupling strength, which scales with the cavity mode volume as $V^{-1/2}$. Nanoscale mode volumes are typically attained in metallic structures – hence the term plasmonic nano- and pico-cavities. Plasmonic cavities are inherently lossy, with Q values as small as ~10. However, the small volumes can compensate for the sub-par Q in Purcell-related effects that scale as ~Q/V, and can even enable strong and ultra-strong light-matter coupling (Section 2C). The broad linewidth of plasmonic cavities may become a virtue in a number of experimental situations. Indeed, an extended linewidth enables interactions with multiple transitions within a single quantum emitter, increasing coupling strength while potentially preserving the single-photon quantum nonlinearities [470]. Plasmonic nanoparticle crystals provide a platform to achieve deep strong light matter coupling, where the light-matter coupling strength exceeds the transition energy of the material[471]. Finaly, scanning tips such as those in s-SNOM or STM apparatus can act as a scanning cavity enabling spectroscopy and imaging in strong light-matter coupling regime [472] down to single molecules[473].

*Extending connections* (cells E-F). Intricate photonic structures have uncovered new paradigms for polaritonic dispersion engineering. Commonly employed structures include waveguide lattices (H) and photonic crystals. Many crystals can support sub-diffractional guided modes, so that surfaces and interfaces can act as natural polaritonic cavities (Fig. 2C and cells F,G). Optical fields in such self-cavities can be confined to length scales ultimately limited by interatomic spacings, giving rise to nano-sculpted polaritons[202] that are naturally amenable to electrostatic/voltage control[182]. Alternatively, (nano)polaritonic properties can be engineered by resorting to QM interfaces, where, e.g., moiré lattices can act as twist-tunable photonic crystals (Fig.5 C-F). Cavities can also induce and modify desired faculties of the light field, such as angular momentum and chirality, opening pathways for emulating the electronic effects in polaritonic platforms. [474,475].

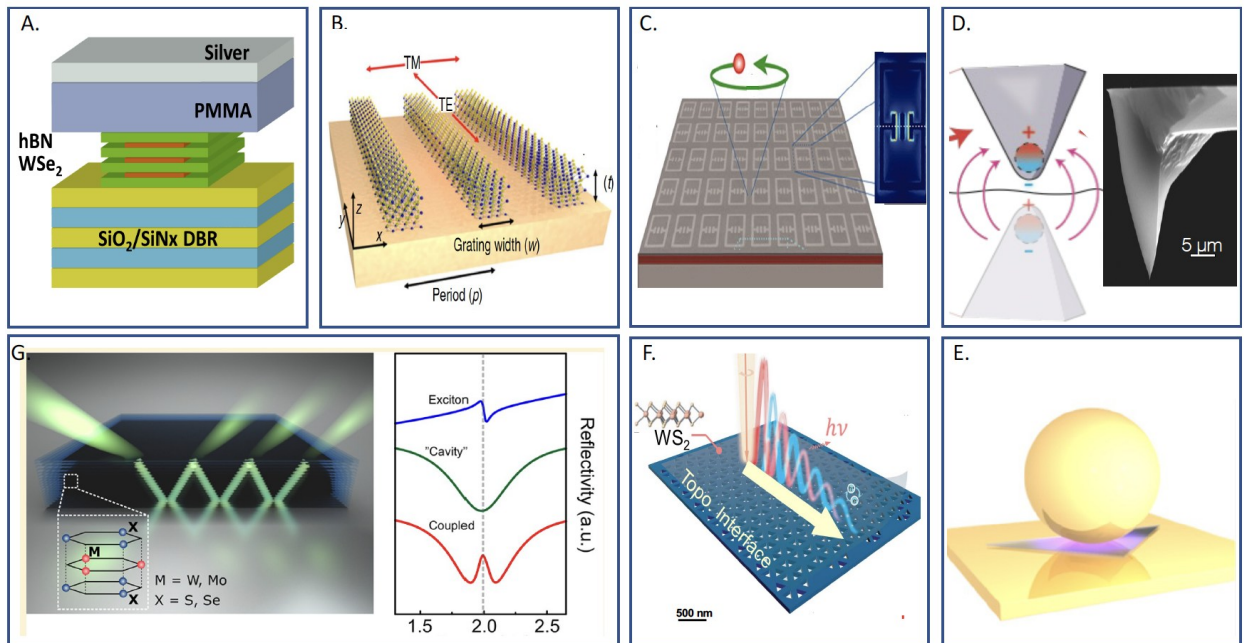

**Box 3: Light management solutions for polaritonic platforms.** Optical cavities (cell A) meta-structures (cells B, C), nano-plasmonic strictures (cells D, E) and photonic/polaritonic crystals (cell F) allows one to localize and enhance light-matter interactions. Cell A: distributed Bragg reflectors (DBRs) are commonly used for engineering polaritonic interactions in solids. Adapted from [456]. Cell B: Schematic of a multi-layer $WS_2$ grating structure on a gold substrate. The coupling of excitons in WS2 with plasmons in AU prompt hybrid plasmon-exciton modes. Red arrows denote the incident light polarization: TM polarization is defined such that the electric field is perpendicular to the grating, while the TE polarization electric field is parallel to the grating. Grate width (w), period (p), and thickness (t) are defined. Adapted from Ref.[476]
Cells C and D: cavity mode volumes can be dramatically reduced by harnessing meta-structures in cell C Refs. [27,149] and even metalized tips of s-SNOM instruments as Cell D. Plasmonic nanocavities enable strong coupling between cyclotron resonances and THz radiation[27,149]. Cell E: plasmonic nano-structures facilitate efficient interfacing with highly confined excitons in atomically layered 2D materials[223]. Cell G: a bulk crystal of van der Waals semiconductors of $MX_2$ is illuminated by a plane wave, which excites a Fabri-Perot cavity mode of the slab (left panel). Reflection spectra at normal incidence modeled for a thin 5 nm $MX_2$ crystal on a glass substrate (blue), from a hypothetical 70 nm $MX_2$ crystal with the oscillator strength of the A-exciton switched off, forming the cavity mode (green), and from a 70 nm TMDC flake with the exciton resonance switched on. Adapted from Ref.[166].

## SECTION 5: What is ahead for polaritonic quantum matter?
Even though it is difficult to predict exactly where all the bits of polaritonic research will land, the contours of a new broad field of research are coming into view.

Polaritonic materials engineering in cavities and nanostructures. The ability to create and control electron-photon states and control their fluctuations in undriven cavities is expected to yield qualitatively new physical phenomena. Thus, engineered dielectric environments (Box 3) can have a profound effect on a material or molecule surrounded by an electromagnetic cavity, even in the absence of external optical drives. This prompts many fundamental questions: Can we consistently increase the conductivity of electrons in materials via hybridization with vacuum fields[477]? Can we destroy a correlated Mott gap to induce a new ground state? Can we induce ferroelectric instability by hybridizing dipole carrying modes of a material with a cavity[208,210]? So far, most attempts to address these vexing questions have been theoretical, but experiments may soon provide direct insights[4]. The BEC of exciton-polaritons (Section 3) and emerging examples of cavity-controlled effects are some of the affirmative experimental examples of the altered ground state in a material-cavity system. The search for material attributes favoring correlated light-matter phases with altered ground states is at the forefront of current research. Especially interesting candidates are moiré superlattices in vdW quantum materials, which have already uncovered a pathway towards systems with ultra-narrow electronic bands[192]. In these tunable materials, the strength of the light-matter interaction can be made comparable to or larger than the electronic kinetic energy, fulfilling preconditions for the polaritonic blockade and other novel states in which electrons and photons are nontrivially entangled. Although the difficulties in attaining superradiant quantum phase transitions mediated by a cavity have been pointed out (based on the no-go theorem[203,478]), the presence of interactions and of higher-order light-matter coupling may lift these constraints.

Many-body physics. Polaritons constitute an ideal platform for studying out-of-equilibrium many-body physics and thermalization in driven-dissipative quantum systems. Examples include the Dicke phase transition, during which the ground state undergoes a transition from a vacuum to a superradiant state with multiple excitations[479,480] and the breakdown of photon blockade caused by increasing drive strength[481]. Naturally, the interplay between loss, pumping, and matter nonlinearities, investigated in atomic, solid and circuit platforms, may open the door to the observation of dissipative quantum phase transitions that are difficult to access in other physical platforms.

Time interfaces. Polaritonic travel in space and time can be controlled with a variety of external stimuli, material interfaces (Sections 2-3) and quantum correlations (Section 4). Notably, time itself is emerging as a capable control resource in photonics[482–484] and one can anticipate that a broader use of time interfaces in polaritonics is on the horizon. Temporal interfaces emerge when material parameters abruptly change in time as a traveling wave of photonic or polaritonic origin traverses across a medium. In close analogy with space interfaces, a temporal interface also prompts back reflections and refraction. In contrast to the properties of a material interface, the frequency of a wave transmitted or refracted through a temporal interface is different from the incident wave[485]. Both the photonic and fluidic implementations of time interfaces are interesting in part because they enable a practical approach for time reversal among many other enigmatic properties. Polaritons and their slow motion through the hosting medium simplify the requirements of an abrupt perturbation that gives rise to time interfaces in the first place.

A unified theory of polaritonic quantum matter. A theory that incorporates all relevant degrees of freedom (photons, electrons, phonons, environment) on an equal footing is needed and wanted. Such a first-principles theory will need to provide a seamless account of entanglement between light and matter[233,277,486]. Theoretically, the description of heavy-photon phenomena requires an understanding of many-body quantum systems with arbitrary environments and interactions. This important theoretical challenge calls for the development of the multiscale and multicomponent QED theoretical framework needed to elucidate the fundamentals of polaritonic-based phenomena and investigate the dichotomy between symmetry-protected topological obstructions and strong interactions.

We conclude that in polaritons, we encounter a concept panoramic in scope, compelling in theoretical foundations, intriguing in technological significance, and captivating in elegance.

**Acknowledgment**

Research polaritons at Columbia is supported as part of Programmable Quantum Materials, an Energy Frontier Research Center funded by the U.S. Department of Energy (DOE), Office of Science, Basic Energy Sciences (BES), under award DE-SC0019443. Research on polaritonic nano-imaging is supported by DE-SC0018426 (PI: DNB). DNB is Moore Investigator in Quantum Materials EPIQS GBMF9455. MKL acknowledges Gordon and Betty Moore Foundation (DOI: 10.37807/gbmf12258) for supporting the research of polaritonic materials.